\documentclass[prd,aps,showpacs,nofootinbib]{revtex4}
\usepackage{amssymb}
\usepackage{graphicx}
\usepackage{amsmath}

\newcommand{\be}{\begin{equation}}
\newcommand{\ee}{\end{equation}}
\newcommand{\bq}{\begin{eqnarray}}
\newcommand{\eq}{\end{eqnarray}}

\begin{document}
\title{An explanation for the tiny value of the cosmological constant and the low vacuum energy density} 
\author{*Cl\'audio Nassif}
\affiliation{\small{*CPFT-Centro de Pesquisas em F\'isica Te\'orica, Rua Rio de Janeiro 1186/s.1304, Lourdes, CEP:30.160-041, 
 Belo Horizonte-MG, Brazil.\\
 cncruz777@yahoo.com.br}}

\begin{abstract}

The paper aims to provide an explanation for the tiny value of the cosmological constant and the low vacuum energy density to
represent the dark energy. To accomplish this, we will search for a fundamental principle of symmetry in space-time by means of the elimination
of the classical idea of rest, by including an invariant minimum limit of speed in the subatomic world. Such a minimum speed, unattainable 
by particles, represents a preferred reference frame associated with a background field that breaks down the Lorentz symmetry. The metric 
of the flat space-time shall include the presence of a uniform vacuum energy density, which leads to a negative
pressure at cosmological length scales. Thus, the equation of state for the cosmological constant [$p$(pressure)$=-\epsilon$
(energy density)] naturally emerges from such a space-time with an energy barrier of a minimum speed. The tiny values of the cosmological 
constant and the vacuum energy density will be successfully obtained, being in agreement with the observational results of Perlmutter,
Schmidt and Riess. 

\end{abstract}

\pacs{98.80.Es; 11.30.Qc\\
DOI:10.1007/s10714-015-1939-8\\
Correlated paper in: http://www.worldscientific.com/worldscinet/ijmpd?journalTabs=read}  

\maketitle

\section{Introduction}

Driven by a search for new fundamental symmetries in Nature\cite{1}, the paper attempts to implement a uniform background field into the
flat space-time. Such a background field connected to a uniform vacuum energy density represents a preferred reference frame, which leads
us to postulate a universal minimum limit of speed for particles with very large wavelengths (very low energies).

The idea that some symmetries of a fundamental theory of quantum gravity may have non trivial consequences for cosmology and particle
physics at very low energies is interesting and indeed quite reasonable. So, it seems that the idea of a universal minimum speed as one 
of the first attempts of Lorentz symmetry violation could have the origin from a fundamental theory of quantum gravity at very low energies
(very large wavelengths).

 The hypothesis of the lowest non-null limit of speed ($V$) for low energies ($v<<c$) in space-time results in the following physical
 reasoning:

- In non-relativistic quantum mechanics, the plane wave wave-function ($Ae^{\pm ipx/\hbar}$) which represents a free particle is an 
idealisation that is impossible to conceive under physical reality. In the event of such an idealized plane wave, it would be possible 
to find with certainty the reference frame that cancels its momentum ($p=0$), so that the uncertainty on its position would be 
$\Delta x=\infty$. However, the presence of an unattainable minimum (non-zero) limit of speed emerges in order to prevent the
ideal case of a plane wave wave-function ($p=constant$ or $\Delta p=0$ with $\Delta x=\infty$). This means that there is no perfect
inertial motion ($v=constant$) such as a plane wave,
except the privileged reference frame of a universal background field connected to an unattainable minimum limit of speed $V$, 
where $p$ would vanish. However, since such a minimum speed $V$ (universal background frame $S_V$) is unattainable for the particles 
with low energies (large wavelengths), their momentum can actually never vanish when one tries to be closer to such a preferred frame
($V$), as it will be shown that there is an insuperable energy barrier when one tries to decelerate a particle very close to the 
vacuum regime of the background frame $S_V$, which represents a fundamental zero-point energy for $v\rightarrow V$ (see Section 5). 

 On the other hand, according to Special Relativity (SR), the momentum cannot be infinite since the maximum speed $c$ is also unattainable
 for a massive particle, except the photon ($v=c$) as it is a massless particle.

 This reasoning allows us to think that the electromagnetic radiation (photon:$``c-c"=c$) as well as the massive particle 
($``v-v">V$ for $v<c$) are in equal-footing in the sense that it is not possible to find a reference frame at rest
 ($v_{relative}=0$) for both through any speed transformation in a space-time with a maximum and a minimum limit of speed. 
Thus such a doubly special relativity with an invariant minimum speed will be denominated as Symmetrical Special Relativity (SSR).
 We will look for new speed transformations of SSR in the next section. 

  In a future paper, we will investigate the origin of the minimum speed $V$, which could have a direct connection with the Planck
length, i.e., the minimum length $l_P=\sqrt{G\hbar/c^3}(\sim 10^{-35}m)$ in a quantum gravity theory. 

 The dynamics of particles in the presence of a universal background reference frame connected to $V$ is within a context of the ideas
 of Sciama\cite{2}, Schr\"{o}dinger\cite{3} and Mach\cite{4}, where there should be an ``absolute" inertial reference frame in relation
 to which we have the inertia of all moving bodies. However, we must emphasize that the approach used here is not classical as machian
 ideas, since the lowest (unattainable) limit of speed $V$ plays the role of a privileged (inertial) reference frame of background field
 instead of the ``inertial" frame of fixed stars.

 It is very curious to notice that the idea of a universal background field was sought in vain by Einstein\cite{5}, motivated firstly 
 by Lorentz. It was Einstein who coined the term {\it ultra-referential} as the fundamental aspect of reality to represent a universal
 background field\cite{6}. Based on such a concept, let us call {\it ultra-referential} $S_V$ to be the universal background field of a
 fundamental inertial reference frame connected to $V$.

 The present doubly special relativity (SSR) is a kind of deformed special relativity (DSR) with two invariant scales, namely the speed 
 of light $c$ and a minimum speed $V$. DSR theory was first investigated by Camelia et al.\cite{7}\cite{8}\cite{9}\cite{10}. It
 contains two invariant scales: the speed of light $c$ and a minimum length scale (Planck length $l_P$ of quantum gravity). An alternative
 approach to DSR theory, inspired by that of Camelia, was proposed later by Smolin and Magueijo\cite{11}\cite{12}\cite{13}.

 Another extension of Special Relativity (SR) is known as Triply Special Relativity, which is characterized by three invariant scales,
 namely the speed of light $c$, a mass $k$ and a length $R$\cite{14}. Still another generalization of SR is the quantizing of
 speeds\cite{15}, where Barrett-Crane spin foam model for quantum gravity with
 positive cosmological constant was considered, encouraging the authors to look for a discrete spectrum of velocities and the physical
 implications of this effect, namely an effective deformed Poincar\'e symmetry.

\section{Transformations of space-time and velocity in the presence of the ultra-referential $S_V$}

 The classical notion we have about the inertial (galilean) reference frames, where the system at rest exists, is eliminated in SSR where
 $v>V(S_V)$ (Fig.1). However, if we consider classical systems composed of macroscopic bodies, the minimum speed $V$ is neglected ($V=0$)
 and so we can reach a vanishing velocity ($v=0$), i.e., in the classical approximation ($V\rightarrow 0$), the
 ultra-referential (background frame) $S_V$ is eliminated and simply replaced by the galilean reference frame $S$ connected 
 to a classical system at rest.

 Since we cannot consider a reference system made up of a set of infinite points at rest in quantum space-time 
 with an invariant minimum speed, then we should define a new status of referentials, namely a non-galilean 
 reference system, which is given essentially as a set of all the particles having the same state of movement (speed $v$) with respect 
 to the ultra-referential $S_V$ (preferred reference frame of the background field), so that $v>V$, $V$ being unapproachable and connected
 to $S_V$. So, a set of particles with the same speed $v$ with respect to the ultra-referential $S_V$ provides a given non-galilean
 framework. Hence, SSR should contain three postulates, namely:

 1)-the non-equivalence (asymmetry) of the non-galilean reference frames due to the presence of the background frame $S_V$ that 
    breaks down Lorentz symmetry, i.e., we cannot exchange $v$ for $-v$ by means of inverse transformations, since we cannot achieve
    a rest state ($v=0$) for a certain non-galilean reference frame in order to reverse the direction of its velocity only for one
    spatial dimension; 

 2)-the invariance of the speed of light ($c$);

 3)-the covariance of the ultra-referential $S_V$ (background framework) connected to an invariant and unattainable minimum limit
    of speed $V$, i.e., all the non-galilean reference frames with speeds $V<v\leq c$ experience the same background frame $S_V$,
   in the sense that the background energy (vacuum energy) at $S_V$ does not produce a flow $-v$ at any of these referentials.
   Thus, $S_V$ does not work like the newtonian absolute space filled by luminiferous (galilean) ether in the old (classical)
   sense, in spite of $S_V$'s being linked to a background energy that works like a {\it non-galilean ``ether''}, leading to the 
   well-known vacuum energy density (cosmological constant), as we will show later. 

 The third postulate is directly connected to the second one. Such a connection will be clarified by investigating the new velocity
 transformations to be obtained soon.

 Of course if we consider $V=0$, we recover the well-known two postulates of SR, i.e., we get the equivalence of inertial reference frames,
 where one can exchange $v$ for $-v$ with appropriate transformations and, consequently, this leads to the absence of such a background 
field ($S_V$); however, the constancy of the speed of light is still preserved.

  Let us assume the reference frame $S^{\prime}$ with a speed $v$ in relation to the ultra-referential $S_V$ according to Fig.1.

\begin{figure}
\includegraphics[scale=0.7]{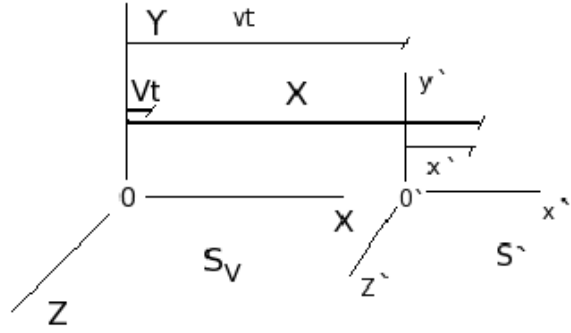}
\caption{$S^{\prime}$ moves with a speed $v(>V)$ with respect to the background field of the covariant ultra-referential $S_V$.
 If $V\rightarrow 0$, $S_V$ is eliminated (empty space) and, thus, the galilean frame $S$ takes place, recovering the Lorentz 
 transformations.}
\end{figure}

So, to simplify, consider the motion only at one spatial dimension, namely $(1+1)D$ space-time with the background field $S_V$.
So we write the following transformations:

  \begin{equation}
 dx^{\prime}=\Psi(dX-\beta_{*}cdt)=\Psi(dX-vdt+Vdt),
  \end{equation}
where $\beta_{*}=\beta\epsilon=\beta(1-\alpha)$, being $\beta=v/c$ and $\alpha=V/v$, so that
$\beta_{*}\rightarrow 0$ for $v\rightarrow V$ or $\alpha\rightarrow 1$.

 \begin{equation}
 dt^{\prime}=\Psi\left(dt-\frac{\beta_{*}dX}{c}\right)=\Psi\left(dt-\frac{vdX}{c^2}+\frac{VdX}{c^2}\right),
  \end{equation}
being $\vec v=v_x{\bf x}$. We have $\Psi=\frac{\sqrt{1-\alpha^2}}{\sqrt{1-\beta^2}}$. If we make
$V\rightarrow 0$ ($\alpha\rightarrow 0$), we recover the Lorentz
transformations, where the ultra-referential $S_V$ is eliminated and simply replaced by the galilean
frame $S$ at rest for a classical observer.

 In order to get the transformations in Eq.(1) and Eq.(2) above, let us consider the following more general transformations:
$x^{\prime}=\theta\gamma(X-\epsilon_1vt)$ and $t^{\prime}=\theta\gamma(t-\frac{\epsilon_2vX}{c^2})$,
 where $\theta$, $\epsilon_1$ and $\epsilon_2$ are factors (functions) to be determined. We hope all these factors
depend on $\alpha$, such that, for $\alpha\rightarrow 0$ ($V\rightarrow 0$), we recover Lorentz transformations as a particular case
 ($\theta=1$, $\epsilon_1=1$ and $\epsilon_2=1$). By using those transformations to perform
$[c^2t^{\prime 2}-x^{\prime 2}]$, we find the identity: $[c^2t^{\prime 2}-x^{\prime 2}]=
\theta^2\gamma^2[c^2t^2-2\epsilon_1vtX+2\epsilon_2vtX-\epsilon_1^2v^2t^2+\frac{\epsilon_2^2v^2X^2}{c^2}-X^2]$.
 Since the metric tensor is diagonal, the crossed terms must vanish and so we assure that
$\epsilon_1=\epsilon_2=\epsilon$. Due to this fact, the crossed terms ($2\epsilon vtX$) are cancelled between
themselves and finally we obtain $[c^2t^{\prime 2}-x^{\prime 2}]=
 \theta^2\gamma^2(1-\frac{\epsilon^2 v^2}{c^2})[c^2t^2-X^2]$. For $\alpha\rightarrow 0$ ($\epsilon=1$ and
$\theta=1$), we reinstate $[c^2t^{\prime 2}-x^{\prime 2}]=[c^2t^2-x^2]$ of SR. Now we write the following
transformations: $x^{\prime}=\theta\gamma(X-\epsilon vt)\equiv\theta\gamma(X-vt+\delta)$ and
$t^{\prime}=\theta\gamma(t-\frac{\epsilon vX}{c^2})\equiv\theta\gamma(t-\frac{vX}{c^2}+\Delta)$, where
we assume $\delta=\delta(V)$ and $\Delta=\Delta(V)$, so that $\delta =\Delta=0$ for $V\rightarrow 0$, which implies $\epsilon=1$.
 So, from such transformations we extract: $-vt+\delta(V)\equiv-\epsilon vt$ and
$-\frac{vX}{c^2}+\Delta(V)\equiv-\frac{\epsilon vX}{c^2}$, from where we obtain
 $\epsilon=(1-\frac{\delta(V)}{vt})=(1-\frac{c^2\Delta(V)}{vX})$. As $\epsilon$ is a dimensionaless factor,
we immediately conclude that $\delta(V)=Vt$ and $\Delta(V)=\frac{VX}{c^2}$, so that we find
$\epsilon=(1-\frac{V}{v})=(1-\alpha)$. On the other hand, we can determine $\theta$ as follows: $\theta$ is a
function of $\alpha$ ($\theta(\alpha)$), such that $\theta=1$ for $\alpha=0$, which also leads to $\epsilon=1$ in
order to recover the Lorentz transformations. So, as $\epsilon$ depends on $\alpha$, we conclude that $\theta$ can
also be expressed in terms of $\epsilon$, namely $\theta=\theta(\epsilon)=\theta[(1-\alpha)]$, where
$\epsilon=(1-\alpha)$. Therefore we can write $\theta=\theta[(1-\alpha)]=[f(\alpha)(1-\alpha)]^k$, where the
exponent $k>0$. The function $f(\alpha)$ and $k$ will be estimated by satisfying the following conditions:

i) as $\theta=1$ for $\alpha=0$ ($V=0$), this implies $f(0)=1$.

ii) the function $\theta\gamma =
\frac{[f(\alpha)(1-\alpha)]^k}{(1-\beta^2)^{\frac{1}{2}}}=\frac{[f(\alpha)(1-\alpha)]^k}
{[(1+\beta)(1-\beta)]^{\frac{1}{2}}}$ should have a symmetrical behavior, that is to say it goes to zero closer
to $V$ ($\alpha\rightarrow 1$) in the same way it goes to infinite closer to $c$ ($\beta\rightarrow 1$). In other words, this
means that the numerator of the function $\theta\gamma$, which depends on $\alpha$ should have the same shape of its denumerator,
  which depends on $\beta$. Due to such conditions, we naturally conclude that $k=1/2$ and
$f(\alpha)=(1+\alpha)$, so that $\theta\gamma=
\frac{[(1+\alpha)(1-\alpha)]^{\frac{1}{2}}}{[(1+\beta)(1-\beta)]^{\frac{1}{2}}}=
\frac{(1-\alpha^2)^{\frac{1}{2}}}{(1-\beta^2)^\frac{1}{2}}=\frac{\sqrt{1-V^2/v^2}}{\sqrt{1-v^2/c^2}}=\Psi$, where
$\theta=\sqrt{1-\alpha^2}=\sqrt{1-V^2/v^2}$.

The transformations shown in Eq.(1) and Eq.(2) are the direct transformations
from $S_V$ [$X^{\mu}=(X,ct)$] to $S^{\prime}$ [$x^{\prime\nu}=(x^{\prime},ct^{\prime})$], where
we have $x^{\prime\nu}=\Lambda^{\nu}_{\mu} X^{\mu}$ ($x^{\prime}=\Lambda X$),
so that we obtain the following matrix of transformation:

\begin{equation}
\displaystyle\Lambda=
\begin{pmatrix}
\Psi & -\beta(1-\alpha)\Psi \\
-\beta(1-\alpha)\Psi & \Psi
\end{pmatrix},
\end{equation}
such that $\Lambda\rightarrow\ L$ (Lorentz matrix of rotation) for $\alpha\rightarrow 0$ ($\Psi\rightarrow\gamma$).

We obtain $det\Lambda=\frac{(1-\alpha^2)}{(1-\beta^2)}[1-\beta^2(1-\alpha)^2]$, where $0<det\Lambda<1$. Since
$V$ ($S_V$) is unattainable ($v>V)$, this assures that $\alpha=V/v<1$ and therefore the matrix $\Lambda$
admits inverse ($det\Lambda\neq 0$ $(>0)$). However, $\Lambda$ is a non-orthogonal matrix
($det\Lambda\neq\pm 1$) and so it does not represent a rotation matrix ($det\Lambda\neq 1$) in such a space-time
due to the presence of the privileged frame of background field $S_V$ that breaks strongly the invariance of the norm of the
4-vector (limit $v\rightarrow V$ in Eq.(30) or Eq.(31)). Actually such an effect ($det\Lambda\approx 0$ for $\alpha\approx 1$
or $v\approx V$) emerges from a new relativistic physics of SSR for treating much lower energies at ultra-infrared regime closer to $S_V$ 
(very large wavelengths).

 We notice that $det\Lambda$ is a function of the speed $v$ with respect to $S_V$. In the approximation for
$v>>V$ ($\alpha\approx 0$), we obtain $det\Lambda\approx 1$ and so we practically reinstate the rotation behavior
of Lorentz matrix $L$ as a particular regime for higher energies. If we make $V\rightarrow 0$ ($\alpha\rightarrow 0$),
we recover $det\Lambda\approx det L=1$ (rotation condition). This subject will be explored with more details in Section 3, 
where we will verify whether Eq.(3) forms a group. We will explore deep physical implications of such a result that will also lead
to the tiny positive value of the cosmological constant (Section 6). 

The inverse transformations (from $S^{\prime}$ to $S_V$) are

 \begin{equation}
 dX=\Psi^{\prime}(dx^{\prime}+\beta_{*}cdt^{\prime})=\Psi^{\prime}(dx^{\prime}+vdt^{\prime}-Vdt^{\prime}),
  \end{equation}

 \begin{equation}
 dt=\Psi^{\prime}\left(dt^{\prime}+\frac{\beta_{*}
 dx^{\prime}}{c}\right)=\Psi^{\prime}\left(dt^{\prime}+\frac{vdx^{\prime}}{c^2}-
\frac{Vdx^{\prime}}{c^2}\right).
  \end{equation}

In matrix form, we have the inverse transformation $X^{\mu}=\Lambda^{\mu}_{\nu} x^{\prime\nu}$
 ($X=\Lambda^{-1}x^{\prime}$), so that the inverse matrix is

\begin{equation}
\displaystyle\Lambda^{-1}=
\begin{pmatrix}
\Psi^{\prime} & \beta(1-\alpha)\Psi^{\prime} \\
 \beta(1-\alpha)\Psi^{\prime} & \Psi^{\prime}
\end{pmatrix},
\end{equation}
where we can show that $\Psi^{\prime}$=$\Psi^{-1}/[1-\beta^2(1-\alpha)^2]$, so that we must satisfy $\Lambda^{-1}\Lambda=I$.

 Indeed we have $\Psi^{\prime}\neq\Psi$ and therefore $\Lambda^{-1}\neq\Lambda(-v)$. This aspect of $\Lambda$ has an important physical
 implication. In order to understand such an implication, let us first consider the rotation aspect of Lorentz matrix in SR. Under SR, we
have $\alpha=0$ ($V=0$), so that $\Psi^{\prime}\rightarrow\gamma^{\prime}=\gamma=(1-\beta^2)^{-1/2}$.
 This symmetry ($\gamma^{\prime}=\gamma$, $L^{-1}=L(-v)$) happens because the galilean reference
frames allow us to exchange the speed $v$ (of $S^{\prime}$) for $-v$ (of $S$) when we are at rest at
$S^{\prime}$. However, under SSR, since there is no rest at $S^{\prime}$, we cannot exchange $v$ (of $S^{\prime}$) for $-v$ (of $S_V$)
due to that asymmetry ($\Psi^{\prime}\neq\Psi$, $\Lambda^{-1}\neq\Lambda(-v)$). Due to this fact,
$S_V$ must be covariant, namely $V$ remains invariant for any change of reference frame in such a space-time. Thus we
can notice that the paradox of twins, which appears due to that symmetry by
exchange of $v$ for $-v$ in SR should be naturally eliminated in SSR, where only the
reference frame $S^{\prime}$ can move with respect to $S_V$. So, $S_V$ remains
covariant (invariant for any change of reference frame). Such a covariance will be verified soon.

  We have $det\Lambda=\Psi^2[1-\beta^2(1-\alpha)^2]\Rightarrow [(det\Lambda)\Psi^{-2}]=[1-\beta^2(1-\alpha)^2]$. So
we can alternatively write $\Psi^{\prime}$=$\Psi^{-1}/[1-\beta^2(1-\alpha)^2]=\Psi^{-1}/[(det\Lambda)\Psi^{-2}]
=\Psi/det\Lambda$. By inserting this result in Eq.(6) to replace $\Psi^{\prime}$, we obtain the relationship
between the inverse matrix $\Lambda^{-1}$ and $\Lambda(-v)$, namely $\Lambda^{-1}=\Lambda(-v)/det\Lambda$. 

By dividing Eq.(1) by Eq.(2), we obtain the following speed transformation:

   \begin{equation}
  v_{rel}=\frac{v^{\prime}-v+V}
{1-\frac{v^{\prime}v}{c^2}+\frac{v^{\prime}V}{c^2}},
   \end{equation}
 where we have considered $v_{rel}=v_{relative}\equiv dx^{\prime}/dt^{\prime}$
 and $v^{\prime}\equiv dX/dt$.  $v^{\prime}$ and $v$ are given with
 respect to $S_V$, and $v_{rel}$ is the relative velocity between $v^{\prime}$ and $v$. Let us consider
 $v^{\prime}>v$ (see Fig.2). 

\begin{figure}
\includegraphics[scale=0.7]{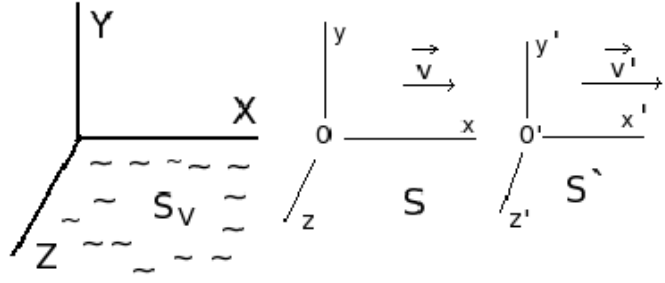}
\caption{$S_V$ is the covariant ultra-referential of background field related to the vacuum energy. $S$ represents the
reference frame for a massive particle with speed $v$ in relation to $S_V$, where $V<v<c$.
  $S^{\prime}$ represents the reference frame for a massive particle with speed $v^{\prime}$
in relation to $S_V$. In this case, we consider $V~(S_V)<v\leq v^{\prime}\leq c$.}
\end{figure}

 If $V\rightarrow 0$, the transformation in Eq.(7) recovers the Lorentz
 velocity transformation where $v^{\prime}$ and $v$ are given in relation to a
 certain galilean frame $S_0$ at rest. Since Eq.(7) implements the ultra-referential
 $S_V$, the speeds $v^{\prime}$ and $v$ are now given with respect to the background frame $S_V$,
  which is covariant (absolute). Such a covariance is verified if we assume that
 $v^{\prime}=v=V$ in Eq.(7). Thus, for this case, we obtain $v_{rel}=``V-V"=V$.

  Let us also consider the following cases in Eq.(7):

 {\bf a)} $v^{\prime}=c$ and $v\leq c\Rightarrow v_{rel}=c$. This
 just verifies the well-known invariance of $c$.

 {\bf b)} if $v^{\prime}>v(=V)\Rightarrow v_{rel}=``v^{\prime}-V"=v^{\prime}$. For
 example, if $v^{\prime}=2V$ and $v=V$ $\Rightarrow v_{rel}=``2V-V"=2V$. This
 means that $V$ really has no influence on the speed of the particles. So $V$ works as if
 it were an ``{\it absolute zero of movement}'', being invariant and having the same value
in all directions of space of the isotropic background field.

 {\bf c)} if $v^{\prime}=v$ $\Rightarrow v_{rel}=``v-v"$($\neq 0)$
$=\frac{V}{1-\frac{v^2}{c^2}(1-\frac{V}{v})}$. From ({\bf c}) let us
consider two specific cases, namely:

  -$c_1$) assuming $v=V\Rightarrow v_{rel}=``V-V"=V$ as verified before.

  -$c_2$) if $v=c\Rightarrow v_{rel}=c$,
 where we have the interval $V\leq v_{rel}\leq c$ for $V\leq v\leq c$.

This last case ({\bf c}) shows us in fact that it is impossible to find the
rest for the particle on its own reference frame $S^{\prime}$, where
$v_{rel}(v)$ ($\equiv\Delta v(v)$) is a function that increases with the increasing of $v$ . However,
 if we make $V\rightarrow 0$, then we would have $v_{rel}\equiv\Delta v=0$ and therefore
it would be possible to find the rest for $S^{\prime}$, which would become simply a galilean
reference frame of SR.

 By dividing Eq.(4) by Eq.(5), we obtain

 \begin{equation}
  v_{rel}=\frac{v^{\prime}+v-V}{1+\frac{v^{\prime}v}{c^2}-\frac{v^{\prime}V}{c^2}}=
 \frac{v^{\prime}+v^*}{1+\frac{v^{\prime}v^*}{c^2}},
   \end{equation}
where we define the notation $v^*=v\epsilon=v(1-\alpha)=v(1-V/v)=v-V$. 

 In Eq.(8), if $v^{\prime}=v=V\Rightarrow ``V+V"=V$. Indeed $V$ is
 invariant, working like an {\it absolute zero state} in SSR. If
 $v^{\prime}=c$ and $v\leq c$, this implies $v_{rel}=c$. For $v^{\prime}>V$
 and considering $v=V$, this leads to $v_{rel}=v^{\prime}$. As a specific example, if $v^{\prime}=2V$
 and assuming $v=V$, we would have $v_{rel} =``2V+V"=2V$. And if
 $v^{\prime}=v\Rightarrow v_{rel}=``v+v"=\frac{2v-V}{1+\frac{v^2}{c^2}(1-\frac{V}{v})}$.
  In newtonian regime ($V<<v<<c$), we recover $v_{rel}=``v+v"=2v$. In relativistic (einsteinian)
regime ($v\rightarrow c$), we reinstate Lorentz transformation for this case ($v^{\prime}=v$),
 i.e., $v_{rel}=``v+v"=2v/(1+v^2/c^2)$.

 By joining both transformations in Eq.(7) and in Eq.(8) into just one, we write the following compact form:

\begin{equation}
  v_{rel}=\frac{v^{\prime}\mp\epsilon v}
{1\mp\frac{v^{\prime}\epsilon v}{c^2}}=\frac{v^{\prime}\mp v(1-\alpha)}
{1\mp\frac{v^{\prime}v(1-\alpha)}{c^2}}=\frac{v^{\prime}\mp v\pm V}
{1\mp \frac{v^{\prime}v}{c^2}\pm \frac{v^{\prime}V}{c^2}},
\end{equation}
being $\alpha=V/v$ and $\epsilon=(1-\alpha)$. For $\alpha=0$ ($V=0$) or $\epsilon=1$, we recover Lorentz speed
transformations.

  Transformations for $(3+1)D$ in SSR will be treated elsewhere. In the next section, we will check whether the new 
transformations given by Eq.(1) and Eq.(2) form a group, giving a physical explanation for such a result.

\section{Do the space-time transformations with an invariant minimum speed form a group? What are their deep implications?}

   It is well-known that the Lorentz transformations form a group ($L=L(v)$), since they obey the following conditions, namely:
a) $L_2L_1=L(v_2)L(v_1)=L(v_3)=L_3\in L(v)$ (Closure condition); b) $L_1(L_2L_3)=(L_1L_2)L_3$ (Associativity); 
c) $L_0L=LL_0=L$, such that $L_0=L(0)=I$ (Identity element); d) $L^{-1}L=LL^{-1}=L_0$, being $L^{-1}=L{(-v)}$ (Inverse element). 

    Our goal is to make an analysis of the new transformations in Eq.(1) and Eq.(2) with regard to the conditions above in order to
verify whether they form a group and discuss deeply the results. So, to do that, we first rewrite the matrix $\Lambda$ (Eq.(3)), namely:

\begin{equation}
\displaystyle\Lambda=
\begin{pmatrix}
\Psi & -\Psi\beta^{*} \\
-\Psi\beta^{*} & \Psi
\end{pmatrix},
\end{equation}
where $\Psi=\frac{\sqrt{1-V^2/v^2}}{\sqrt{1-v^2/c^2}}$. We have defined the notation $\beta^*=\beta\epsilon=\beta(1-\alpha)=
(v/c)[1-V/v]$. If $V\rightarrow 0$ or $\alpha\rightarrow 0$, we recover the Lorentz matrix, i.e., $\Lambda(v)\rightarrow L(v)$,
since $\Psi\rightarrow\gamma$ and $\beta^*\rightarrow\beta$. 

  Now, we have $\Lambda_1=\Lambda(v_1)$ as being 

\begin{equation}
\displaystyle\Lambda_1=
\begin{pmatrix}
\Psi_1 & -\Psi_1\beta_1^{*} \\
-\Psi_1\beta_1^{*} & \Psi_1
\end{pmatrix}= 
\begin{pmatrix}
\Psi_1 & -\Psi_1\frac{v_1^*}{c} \\
-\Psi_1\frac{v_1^*}{c} & \Psi_1
\end{pmatrix}
\end{equation}

  and $\Lambda_2=\Lambda(v_2)$ as being

\begin{equation}
\displaystyle\Lambda_2=
\begin{pmatrix}
\Psi_2 & -\Psi_2\beta_2^* \\
-\Psi_2\beta_2^* & \Psi_2
\end{pmatrix}=
\begin{pmatrix}
\Psi_2 & -\Psi_2\frac{v_2^*}{c} \\
-\Psi_2\frac{v_2^*}{c} & \Psi_2
\end{pmatrix}, 
\end{equation}
so that $\Lambda_2\Lambda_1$ is 

\begin{equation}
\displaystyle\Lambda_2\Lambda_1=[\Psi_2\Psi_1(1+\beta^*_2\beta^*_1)]
\begin{pmatrix}
1 & -\frac{(\beta_1^*+\beta_2^*)}{1+\beta_2^*\beta_1^*} \\
-\frac{(\beta_1^*+\beta_2^*)}{1+\beta_2^*\beta_1^*} & 1
\end{pmatrix},
\end{equation}
where $\beta_1^*=\beta_1\epsilon_1=\beta_1(1-\alpha_1)=(v_1/c)[1-V/v_1]$ and $\beta_2^*=\beta_2\epsilon_2=\beta_2(1-\alpha_2)=
(v_2/c)[1-V/v_2]$.

We obtain that the multiplicative term of the matrix in Eq.(13) is written as $\Psi_2\Psi_1(1+\beta^*_2\beta^*_1)=
\sqrt{(1-V^2/v_2^2)(1-V^2/v_1^2)}\frac{1+(v_1^*v_2^*/c^2)}{\sqrt{1-(v_1^2/c^2+v_2^2/c^2-v_1^2v_2^2/c^4)}}$. Now by inserting 
this term into Eq.(13), we rewrite Eq.(13) in the following way: 

\begin{equation}
\displaystyle\Lambda_2\Lambda_1=\frac{\sqrt{\left(1-\frac{V^2}{v_2^2}\right)\left(1-\frac{V^2}{v_1^2}\right)}
\left(1+\frac{v_1^*v_2^*}{c^2}\right)}{\sqrt{1-\left(\frac{v_1^2}{c^2}+\frac{v_2^2}{c^2}-\frac{v_1^2v_2^2}{c^4}\right)}}
\begin{pmatrix}
1 & -\frac{1}{c}\left(\frac{v_1^*+v_2^*}{1+ \frac{v_1^*v_2^*}{c^2}}\right) \\
-\frac{1}{c}\left(\frac{v_1^*+v_2^*}{1+\frac{v_1^*v_2^*}{c^2}}\right) & 1
\end{pmatrix}
\end{equation}

Now we should note that, if the Eq.(14) satisfies the closure condition, Eq.(14) must be equivalent to 

\begin{equation}
\displaystyle\Lambda_2\Lambda_1=\Lambda_3=\Psi_3
\begin{pmatrix}
1  & -\frac{v_3^*}{c} \\
-\frac{v_3^*}{c} & 1
\end{pmatrix}, 
\end{equation}
 where, by comparing Eq.(14) with Eq.(15), we must verify whether the closure condition is satisfied, i.e., $\Psi_3\equiv\sqrt{(1-V^2/v_2^2)(1-V^2/v_1^2)}\frac{1+(v_1^*v_2^*/c^2)}
{\sqrt{1-(v_1^2/c^2+v_2^2/c^2-v_1^2v_2^2/c^4)}}$ and  $v_3^*\equiv(v_2^*+v_1^*)/[1+(v_2^*v_1^*)/c^2]$. However, we first realize that such
 speed transformation, which should be obeyed in order to satisfy the closure condition, differs from the correct speed transformation 
(Eq.(8)) that has origin from the space-time transformations with a minimum speed given in Eq.(4) and Eq.(5). So, according to Fig.2, 
if we simply redefine $v^{\prime}=v_2$ and $v=v_1$, we rewrite the correct transformation (Eq.(8)) as being
$v_{rel}=v_3=(v_2+v_1^*)/[1+(v_2v_1^*)/c^2]$ with $v_1^*=v_1-V$. Now, we see that the correct transformation for $v_3$ (Eq.(8)) is not the
same transformation given in the matrix above (Eq.(14)), i.e., we have $v_3\neq (v_2^*+v_1^*)/[1+(v_2^*v_1^*)/c^2]$. 

One of the conditions for having the closure relation is that the components outside the diagonal of the matrix in Eq.(14) or Eq.(15) must include
$v_3$ given by Eq.(8), which does not occur. Therefore, we are already able to conclude that such condition is not obeyed in a spacetime
with a minimum speed (a preferred reference frame) at the subatomic level, i.e., we find $\Lambda_2\Lambda_1\neq\Lambda_3$, which does not 
generate a group. In order to clarify further this question, we just make the approximation $V=0$ or also $v_1\gg V$ and $v_2\gg V$ in
Eq.(14), and thus we recover the closure relation of the Lorentz group, as follows: 

\begin{equation}
\displaystyle(\Lambda_2\Lambda_1)_{V=0}=L_2L_1=\frac{\left(1+\frac{v_1v_2}{c^2}\right)}
{\sqrt{1-\left(\frac{v_1^2}{c^2}+\frac{v_2^2}{c^2}-\frac{v_1^2v_2^2}{c^4}\right)}}
\begin{pmatrix}
1 & -\frac{1}{c}\left(\frac{v_1+v_2}{1+\frac{v_1v_2}{c^2}}\right) \\
-\frac{1}{c}\left(\frac{v_1+v_2}{1+\frac{v_1v_2}{c^2}}\right) & 1
\end{pmatrix}=L_3, 
\end{equation}
where

\begin{equation}
\displaystyle L_3=\gamma_3
\begin{pmatrix}
 1  &     -\frac{v_3}{c}\\
 -\frac{v_3}{c}  & 1
\end{pmatrix}, 
\end{equation}
which is the closure condition of the Lorentz group, since now it is obvious that the Lorentz transformation of speeds appears outside 
the diagonal of the matrix in Eq.(16), i.e., we find $v_3=(v_1+v_2)/[1+(v_1v_2)/c^2]$. And, in order to complete the verification of the
closure condition above, it is easy to verify that the multiplicative term of the matrix (Eq.(16)) is $\gamma_2\gamma_1(1+\beta_2\beta_1)=
\left(1+\frac{v_1v_2}{c^2}\right)/\sqrt{1-\left(\frac{v_1^2}{c^2}+\frac{v_2^2}{c^2}-\frac{v_1^2v_2^2}{c^4}\right)}=\gamma_3$. To do this,
we have to consider $v_3=(v_1+v_2)/[1+(v_1v_2)/c^2]$, so that we use this transformation to be inserted into $\gamma_3=1/\sqrt{1-v_3^2/c^2}$ and
we finally show that $\gamma_3=1/\sqrt{1-v_3^2/c^2}=\gamma_2\gamma_1(1+\beta_2\beta_1)$. However, now starting from this same procedure
for obtaining $\Psi_3=\frac{\sqrt{1-V^2/v_3^2}}{\sqrt{1-v_3^2/c^2}}$, where we have to use the correct transformation for $v_3$ (Eq.(8)),
we verify that $\Psi_3\neq\Psi_2\Psi_1(1+\beta_2^*\beta_1^*)$ and thus we conclude definitively that the closure condition does not apply to
the new transformations, i.e., indeed we have $\Lambda_2\Lambda_1\neq\Lambda_3$. 

Although we already know that the new transformations do not form a group, it is still important to provide a physical justification 
for such conclusion. To do this with more clarity, we also should investigate whether the identity element and the inverse element 
exist in such a spacetime with an invariant minimum speed, since these two conditions are relevant to give us a clear comprehension
of the conception of motion in this spacetime. 

 {\bf(I) Identity element} \\

For the case $(1+1)D$, the Lorentz group provides the identity element $L_0=I_{(2X2)}$, since $L_0L=LL_0=L$. As the Lorentz matrix 
is $\displaystyle L_{(2X2)}=\begin{pmatrix}
 \gamma  &     -\beta\gamma \\
-\beta\gamma  & \gamma
\end{pmatrix}$, it is easy to see that, if we make $v=0$ or $\beta=0$ (rest condition), the Lorentz matrix recover the identity matrix $\displaystyle
 I_{(2X2)}=\begin{pmatrix}
 1 &     0 \\
 0  &    1
\end{pmatrix}$, since $\gamma_0=\gamma(v=0)=1$. This trivial condition of rest plus the fact that $det(L)=1$ (rotation matrix $L$)
 shows us the indistinguishability of rest and inertial motion. 

The new transformations are represented by the matrix $\displaystyle\Lambda=
\begin{pmatrix}
\Psi & -\beta(1-\alpha)\Psi \\
-\beta(1-\alpha)\Psi & \Psi
\end{pmatrix}$, where we have $\beta^*=\beta\epsilon=\beta(1-\alpha)$, with $\alpha=V/v$. Now, it is important to notice that there is no any speed $v$
that generates the identity matrix from the new matrix. We would expect that the hypothesis $v=V$ could do that, but,
if we make $v=V$ ($\alpha=1$) inside the new matrix, we find the null matrix, i.e., $\displaystyle\Lambda(V)=\begin{pmatrix}
0 & 0\\
0 & 0
\end{pmatrix}$, since $\Psi(V)=0$. So, we obtain $\Lambda(V)\Lambda=\Lambda_V\Lambda=\Lambda\Lambda_V=\displaystyle\begin{pmatrix}
0 & 0\\
0 & 0
\end{pmatrix}\neq\Lambda$, where we have $\Lambda=\Lambda(v>V)$. Thus, there is no identity element in this spacetime, which means that
there should be a distinction of motion and rest, since there is a preferred reference frame (an invariant minimum speed)  
in respect to which, the motion $v(>V)$ is given, in view of the absence of the rest condition for particles in this spacetime.\\

{\bf(II) Inverse element} \\

It is well-known that the inverse element exists in Lorentz transformations that form a group, i.e., we have $L^{-1}(v)=L(-v)$, which means
that we can exchange the observer in the reference frame $S$ at rest by another observer in the reference frame $S^{\prime}$ with speed $v$ in 
respect to $S$, so that the other observer at $S^{\prime}$ simply observes $S$ with a speed $-v$. Such symmetry comes from the galilean
relativity of motion, which it is essentially due to the indistinguishability of rest and inertial motion. Here we must stress that such
indistinguishability is broken down in the new transformations, since the invariant minimum speed related to a background reference
frame introduces a preferential motion $v(>V)$ that cannot be exchanged by $-v$ due to the distinction of motion and rest, since rest
does not exist in this spacetime, where we get $\Lambda^{-1}(v)\neq\Lambda(-v)$, such that we obtain
 $\Lambda(-v)\Lambda(v)=\displaystyle\theta^2
\begin{pmatrix}
\gamma & \beta(1-\alpha)\gamma \\
\beta(1-\alpha)\gamma & \gamma
\end{pmatrix}$ $\times$
$\displaystyle\begin{pmatrix}
\gamma & -\beta(1-\alpha)\gamma \\
-\beta(1-\alpha)\gamma & \gamma
\end{pmatrix}=\displaystyle\Psi^2\begin{pmatrix}
 \left(1-\frac{v^{*2}}{c^2}\right) & 0 \\
 0 & \left(1-\frac{v^{*2}}{c^2}\right)
\end{pmatrix}\neq I_{(2X2)}$. For $V=0$ ($\alpha=0$), we recover the inverse element of the Lorentz group, that is a rotation group.\\

In short, we have verified that the new transformations do not form a group and we have provided a physical explanation for such Lorentz
violation in view of the existence of an invariant minimum speed that breaks down the indistinguishability of rest and motion. 

We have also concluded that the new matrix $\Lambda$ (Eq.(3)) does not represent a rotation matrix (Section 2). In view of this,
we can realize that such transformations are not related with the well-known rotation group $SO(3)$ (Lie group), whose elements
$R(\vec\alpha)$ and $R(\vec\beta)$ should obey a closure condition $R(\vec\alpha)R(\vec\beta)=R(\vec\gamma)$, such that $\vec\gamma=
\gamma(\vec\alpha,\vec\beta)$, with $R(\vec\gamma)$ being a new rotation that belongs to the group, so that
$det(R)=+1$ (rotation condition), while we find $det(\Lambda)=\theta^2\gamma^2\left[1-\frac{v^2(1-\alpha)^2}{c^2}\right]$,
 where $0<det(\Lambda)<1$, violating the rotation condition. 

Although there could be a more complex mathematical structure in order to encompass the new transformations, which should be deeply
investigated, at least, here we will make some interesting mathematical approximations on $\Lambda$ in order to help us to understand
further the nature of the new transformations. 

In a certain approximation, let us show that $\Lambda$ is a combination of rotation and deformation of the space-time interval $ds$, 
reminding the polar decomposition theorem in linear algebra for a ``rigid" body that rotates and deforms. Intuitively, the polar 
decomposition separates a certain matrix $A$ into a component that stretches the space along a set of orthogonal axes, represented by $P$,
and a rotation (with possible reflection) represented by $U$, i.e., $A=UP$. where $U$ is a unitary matrix and $P$ is a Hermitian matrix.

When a rigid body rotates, its length remains invariant. This effect is analogous to the invariance of the space-time interval $ds$ under
the Lorentz transformation $L$ (Lorentz group) due to a rotation.

When a ``rigid'' body deforms, such effect is analogous to a deformation (e.g: stretching) of the spacetime interval $ds$ that occurs close 
to the minimum speed (see Eq.(30) and Eq.(31) for $v\rightarrow V$). Thus, at a first sight, the new transformation $\Lambda$ could be
written simply as a polar decomposition, so that $\Lambda=LD$, where $L$ is a rotation matrix (Lorentz matrix) as a special case of the
unitary matrix $U$ and $D$ is a deformation matrix (symmetric matrix) as a special case of the Hermitian matrix $P$, since Hermitian 
matrices can be understood as the complex extension of real symmetric matrices. However, we can verify that such analogy fails 
quantitatively when one tries to calculate an exact matrix $D_{(2x2)}$ that satisfies the linear decomposition $LD=\Lambda$ for any 
speed $v$ (any energy scale), i.e., there is no $D_{(2x2)}$ that satisfies such decomposition, because it seems that we have
a kind of non-linear or inseparable combination of rotation and deformation. So, in order to accomplish a stronger analogy with
the polar decomposition, we need to make some mathematical approximations on the matrix $\Lambda$ in such a way that we can be able
to separate both effects of rotation and deformation. To do this, let us first write the matrix $\Lambda$ in the following way:   

\begin{equation}
\displaystyle\Lambda=\frac{\sqrt{1-\alpha^2}}{\sqrt{1-\beta^2}}
\begin{pmatrix}
1  & -\frac{v(1-\alpha)}{c} \\
-\frac{v(1-\alpha)}{c} & 1
\end{pmatrix}
\end{equation}

Instead of making $\alpha=0$ (or $V=0$) in order to recover the Lorentz matrix $L$, here we make an alternative approximation, namely
$v>>V$, which means that, for higher energies, we recover practically the matrix $L$ (rotation). On the other hand, for much lower 
energies, i.e., for $\alpha\approx 1$ (or $v\approx V$), we get 

\begin{equation}
\displaystyle\Lambda_{(v\approx V)}\approx\sqrt{1-\alpha^2}
\begin{pmatrix}
1  & 0 \\
0  & 1
\end{pmatrix}=
\displaystyle\theta
\begin{pmatrix}
1  & 0 \\
0  & 1
\end{pmatrix},
\end{equation}
where we have considered the L'H\^opital's rule, by calculating
 $lim_{\alpha\rightarrow 1}\frac{(1-\alpha)}{\sqrt{1-\alpha^2}}=lim_{\alpha\rightarrow 1}\frac{\sqrt{1-\alpha^2}}{\alpha}=0$. In other
words, this means that $\sqrt{1-\alpha^2}|_{\alpha\approx 1}>>(1-\alpha)|_{\alpha\approx 1}$, so that we can neglect $(1-\alpha)$ with
respect to $\sqrt{1-\alpha^2}$ in such an approximation ($v\approx V$), i.e, we make $\epsilon=(1-\alpha)=0$ into $\Lambda$,
keeping the factor $\theta=\sqrt{1-\alpha^2}$. We obtain $det[\Lambda_{(v\approx V)}]=\theta^2=(1-\alpha^2)=(1-V^2/v^2)$. 

The inverse matrix $\Lambda^{-1}_{(v\approx V)}$ is

\begin{equation}
\displaystyle\Lambda_{(v\approx V)}^{-1}\approx\frac{1}{\sqrt{1-\alpha^2}}
\begin{pmatrix}
1  & 0 \\
0  & 1
\end{pmatrix}=
\displaystyle\theta^{-1}
\begin{pmatrix}
1  & 0 \\
0  & 1
\end{pmatrix},
\end{equation}
where $det[\Lambda_{(v\approx V)}^{-1}]=\theta^{-2}=(1-\alpha^2)^{-1}=(1-V^2/v^2)^{-1}$. 

Both the symmetric matrices in Eq.(19) and Eq.(20) represent deformations, where, for instance, the matrix in Eq.(20) 
leads to a stretching of the space-time interval $ds$ when $v$ is closer to $V$. We realize that such deformations given only 
for much lower energies close to the background frame $S_V$, i.e., the matrices $\Lambda_{v\approx V}=\theta I$ and 
$\Lambda_{v\approx V}^{-1}=\theta^{-1}I$ do not belong to the structure of the Lie group connected to the indentity matrix $I$. 

We conclude that the matrix $\Lambda$ in Eq.(18) already contains effects of deformation ($ds^{\prime}\neq ds$), which become completely evident
for much lower energies ($v\approx V$), where $det(\Lambda)\approx 0$, but, when the speed $v$ increases drastically, i.e., $v>>V$, so, now,
the rotations of Lorentz group are pratically recovered ($det(\Lambda)\approx det(L)=1$) and, thus, we recover the invariance $ds^{\prime}=ds$. With
such approximations, the polar decomposition is practically valid by making $D=\Lambda_{(v\approx V)}$, so that we can verify the
product $LD=L\Lambda_{(v\approx V)}=L(\theta I)\approx\Lambda$, where $\Lambda_{(v\approx V)}$ is a symmetric matrix, which is exactly
the reason for the effects of deformation of $ds$ close to $V$.   

We finally conclude that the set formed by the matrices that appear above does not have a group structure or cannot be considered as a
Poincar\'e's subgroup. This point must be discussed in depth.

Our next step will be to make an investigation of the main effect obtained directly from the violation of the rotation structure at
much lower speeds ($v\approx V$). Such an effect should naturally lead to other deep implications, which will be pointed out, so that 
we will realize that the whole theory contains elements that are connected by a same mathematical and physical structure.

When we make a Lorentz transformation $L(v)$ from the frame $S(v=0)$ to $S^{\prime}$ with speed $v$ with regard to $S$, we have the
well-known ``boost". As the boosts represent rotations, the minimal boost is the identity matrix $L(v=0)=L(0)=I$ connected to the rest
state, such that $L(0)X=X$. However, as such minimal boost does not make sense in this spacetime with a minimum speed that prevents the
rest state, we must stress that the component $\Lambda_{(v\approx V)}(=\theta I)$ in the new transformation 
($\Lambda \approx L\Lambda_{(v\approx V)})$ leads to a non-existence of boosts only in the approximation for much lower energies
($v\rightarrow V$ or  $\alpha\rightarrow 1$), due to the fact that we get $\theta=\sqrt{1-\alpha^2}<<1$. So, only for higher energies 
($v>>V$ or $\alpha\approx 0$), we get $\theta\approx 1$ and, thus, $\Lambda_{(v\approx V)}=\theta I\approx I$, recovering the regime
where the boosts take place (Lorentz group). 

In short, the effects of ``boosts" are generally weakened in this spacetime, mainly in the regime when $v\approx V$, i.e., much closer to the background 
frame $S_V$. So, it is important to stress that, in such special regime, there are no boosts and, therefore, the transformation 
$\theta I$ has another meaning since it does not lead to the change of reference frames. We will go deeper into this issue.

Actually, the symmetric matrix $\theta I$ (Eq.(19)) is the reason of breaking the structure of rotation group (boosts) and it should be interpreted
just as a scale transformation ($\theta$) that provides a variation of the usual space-time interval ($ds$) in function of speed $v$, 
especially when $v$ is close to $V$. The deep physical implication directly related to the effect of variation of $ds$ will be
investigated below. 

Since the matrix $\theta I$ just deforms the interval $ds$, this transformation does not act for changing the reference frames. 
In view of this, we use the following notation to represent such a scale transformation, namely: 

\begin{equation}
 x^{*\mu}=\theta I x^{\mu}, 
\end{equation}
where $x^{*\mu}$ is the deformed vector, $\theta$ being a scale factor, since $\Lambda_{(v\approx V)}x^{\mu}=\theta x^{\mu}=x^{*\mu}$, 
so that we get

 \begin{equation}
ds^{*2}=ds^2(v)=dx^{*\mu}dx^*_{\mu}=\theta^{2}ds^{2}=det[\Lambda_{(v\approx V)}]ds^2, 
 \end{equation}
where $ds^{2}=dx^{\mu}dx_{\mu}$ is the usual squared space-time interval of SR and $ds^{*2}$ is the deformed squared space-time interval
due to new relativistic effects closer to V (no boosts). 

As the usual interval $ds$ does not remain invariant in this spacetime, specially when $v\approx V$; so according to Eq.(22), we realize
that the deformed interval $ds^{*}$ should be the new invariant interval under the change of reference frames in this flat spacetime
with the presence of the background frame $S_V$, such that $ds^{*\prime}=ds^{*}$. In doing this, we introduce a new invariance of deformed
intervals in SSR, namely: 

\begin{equation}
ds^{*\prime 2}=ds^{*2}=g_{\mu\nu}dx^{\mu}dx^{\nu}, 
\end{equation}
where $ds^{*2}=(1-\alpha^2)dx^{\mu}dx_{\mu}=\theta^2 ds^2$ and $ds^{*\prime 2}=(1-\alpha^2)dx^{\prime\mu}dx^{\prime}_{\mu}=\theta^2
ds^{\prime 2}$. Of course if we make $V\rightarrow 0$ or $\alpha\rightarrow 0$ in Eq.(23), we recover the invariance of the usual 
(non-deformed) $ds$ of SR, i.e., $ds^{\prime 2}=ds^{2}=g_{\mu\nu}dx^{\mu}dx^{\nu}$. 

Indeed we realize that the deformed interval $ds^{*}=\theta ds=\sqrt{1-V^2/v^2}ds$ remains finite (Eq.23), since, in the limit 
of $\theta\rightarrow 0$ ($v\rightarrow V$), the usual interval $ds$ undergoes a very large stretching, i.e., $ds\rightarrow\infty$.

From Eq.(23), we obtain 

\begin{equation}
ds^{*\prime}=ds^{*}=\sqrt{g_{\mu\nu}dx^{\mu}dx^{\nu}}=\sqrt{c^2dt^2-dx^2}, 
\end{equation}
where, in this case, we have $dy=dz=0$ and $ds^{*\prime}=\sqrt{(1-\alpha^2)dx^{\prime\mu}dx^{\prime}_{\mu}}=\sqrt{(1-\alpha^2)(c^2dt^{\prime2}-dx^{\prime2})}$. Now, if
we make $dx^{\prime}=0$ (or $x^{\prime}=0$, i.e., at the origin of the reference frame $S^{\prime}$), from Eq.(24) we obtain 

\begin{equation}
ds^{*\prime}=cd\tau^{*}=\sqrt{1-\alpha^2}cd\tau=\sqrt{1-\alpha^2}cdt^{\prime}=\sqrt{c^2dt^2-dx^2}, 
\end{equation}
where we have considered $dt^{\prime}=d\tau$ and so $dt^{\prime *}=d\tau^{*}$, with $d\tau^{*}(=ds^{*\prime}/c=\sqrt{g_{\mu\nu}dx^{\mu}dx^{\nu}}/c)$ being the deformed 
proper time interval, where we have $d\tau^{*}=\theta d\tau=\sqrt{1-\alpha^2}d\tau$. This result has a deep physical implication that has origin in the
breakdown of the structure of Lorentz group.

 Now we are ready to investigate the physical implication from Eq.(25). So, by simply making $dx=vdt$ in Eq.(25) and performing the
calculations, we finally obtain

\begin{equation}
 d\tau\sqrt{1-\alpha^2}=dt\sqrt{1-\beta^2}
\end{equation}

and, then

\begin{equation}
 \Delta\tau\sqrt{1-\frac{V^2}{v^2}}=\Delta t\sqrt{1-\frac{v^2}{c^2}},
\end{equation}
where $\Delta\tau$ is the proper time interval and $\Delta t$ is the improper one. Eq.(27) is the immediate physical implication of the violation of Lorentz group by means of the symmetric matrix $\theta I$ (Eq.(19)) that deforms the proper time, so that 
we can also write Eq.(27) as being $\sqrt{det(\theta I)}\Delta\tau=\theta\Delta\tau=\Delta t\sqrt{1-v^2/c^2}$, where $\theta=\sqrt{1-V^2/v^2}$. 

It is important to call attention to the fact that Eq.(27) shows us that the proper time interval $\Delta\tau$ depends on speed $v$ and, thus, now
it can also be deformed (dilated) like the improper time interval. So, we realize that Eq.(27) reveals a perfect symmetry in the sense that both
intervals of time $\Delta t$ and $\Delta\tau$ can dilate, namely $\Delta t$ dilates for $v\rightarrow c$ and, on the other hand,
$\Delta\tau$ dilates for $v\rightarrow V$. But, if we make $V\rightarrow 0$, we break down such new symmetry of SSR and so we recover the 
well-known time equation of SR, where only $\Delta t$ dilates and $\Delta\tau$ remains invariant.

From Eq.(27) we notice that, if we make $v=v_0=\sqrt{cV}$ (a geometric average between $c$ and $V$), we find exactly the equality 
$\Delta\tau$ (at $S^{\prime}$)=$\Delta t$ (at $S$), namely this is a newtonian result where the time intervals are the same. Thus we conclude 
that $v_0$ represents a special intermediary speed in SSR ($V<<v_0<<c$) such that, if: 

a) $v>>v_0$ (or $v\rightarrow c$), we get $\Delta\tau<<\Delta t$. This is the well-known {\it improper time dilation}.

b) $v<<v_0$ (or $v\rightarrow V$), we get $\Delta\tau>>\Delta t$. Let us call such a new effect as {\it improper time contraction} or
{\it dilation of the proper time interval $\Delta\tau$ with respect to the improper time interval $\Delta t$}. This new effect becomes
more evident only for $v~(S^{\prime})\rightarrow V~(S_V)$, so that, in this limit, we have $\Delta\tau\rightarrow\infty$ for a certain $\Delta t$ fixed 
as being finite. In other words, this means that the proper time ($S^{\prime}$) can now elapse much faster than the improper one.

 It is interesting to notice that we restore the newtonian regime when $V<<v<<c$, which represents a regime of intermediary speeds, 
 so that we get the newtonian approximation from Eq.(27), i.e., $\Delta\tau\approx\Delta t$.

 Squaring both members of Eq.(27) ($\Delta t=\Psi\Delta\tau=\theta\gamma\Delta\tau$) and manipulating the result, we write Eq.(27) as
follows: 
 
   \begin{equation}
    c^2\Delta\tau^2=\frac{1}{(1-\frac{V^2}{v^2})}[c^2\Delta t^2-v^2\Delta t^2]
   \end{equation}

 By placing Eq.(28) in a differential form and manipulating it, we obtain

   \begin{equation}
   c^2\left(1-\frac{V^2}{v^2}\right)\frac{d\tau^2}{dt^2} + v^2=c^2
   \end{equation}

 Eq.(29) shows us that both speeds related to the marching of time  (``temporal-speed''$v_t=c\sqrt{1-V^2/v^2}d\tau/dt$)
 and the spatial speed $v$ form the vertical and horizontal legs of a rectangular triangle respectively (Fig.3). The hypotenuse of the
 triangle is $c=(v_t^2+v^2)^{1/2}$, which represents the spatio-temporal speed of any particle. If $V\rightarrow 0$ in Eq.(29), we recover
 the time equation in SR, i.e., $c^2(d\tau^2/dt^2)+v^2=c^2$. 

Looking at Fig.3, now we see clearly three important cases, namely:

 a) If $v\approx c$, $v_t\approx 0$ (the marching of proper time in $S^{\prime}$ is much slower than in $S$), such that 
$\Delta t>>\Delta\tau$, with $\Psi\approx\gamma>>1$, leading to the well-known dilation of the improper time. 

 b) If $v=v_0=\sqrt{cV}$, $v_t=\sqrt{c^2-v_0^2}$, i.e., the marching of time in $S^{\prime}$ is faster, but it is still 
 in an intermediary regime, such that $\Delta t=\Delta\tau$, with $\Psi=\Psi_0=\Psi(v_0)=1$ (newtonian regime). 

 c) If $v\approx V(<<v_0)$, $v_t\approx\sqrt{c^2-V^2}=c\sqrt{1-V^2/c^2}$ (the marching of proper time is even faster), such that 
$\Delta t<<\Delta\tau$, with $\Psi\approx\theta<<1$ (dilation of the proper time). To illustrate this new effect of proper time dilation,
 let us consider a box that contains an ideal gas with $N$ particles in the frame $S$ of a laboratory. Since the minimum speed 
 $V$ has a microscopic origin, then by considering an average speed per particle of the gas (atom or molecule), we should have such average
 speed $v_{rms}$ ($S^{\prime}$) close to $V$ ($S_V$) only when the temperature of the gas is $T\rightarrow 0$K 
($v_{rms}=\sqrt{\left<v\right>^{2}_N}\rightarrow V$). Thus, an imaginary clock in thermal equilibrium with such ultra-cold system should measure a dilated time interval with 
respect to the time interval measured in the observer's clock (laboratory $S$), or in other words, we could say that ultra-cold systems
 ``grow old" more rapidly, contrary to higher energies when one grows old more slowly. A great experimental effort should be made in order
 to detect the effect of proper time dilation, but, before this, we must search for the origin of the invariant minimum speed. 

\begin{figure}
\includegraphics[scale=0.8]{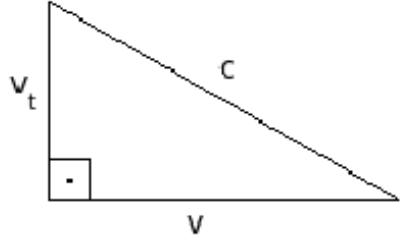}
\caption{We see that the horizontal leg represents the spatial-speed $v$, while the vertical leg represents the temporal-speed $v_t$
(marching of time), where $v_t=\sqrt{c^2-v^2}=c\sqrt{1-v^2/c^2}=c\sqrt{1-V^2/v^2}d\tau/dt$ (see Eq.(27)), so that we always have 
$v^2+v_t^2=c^2$. In SR, when $v=0$, the horizontal leg vanishes (no spatial speed) and so the vertical leg becomes maximum
($v_t=v_{tmax}=c$). However, now according to SSR, due to the existence of a minimum limit of spatial speed ($V$), we can never nullify
the horizontal leg, so that the maximum temporal speed (maximum vertical leg) is $v_{tmax}=\sqrt{c^2-V^2}=c\sqrt{1-V^2/c^2}<c$. On the 
other hand $v_t$ (the vertical leg) cannot be zero since $v=c$ is forbidden for massive particles. So we conclude that the rectangular
triangle is always preserved since both temporal and spatial speeds cannot vanish and, thus, they always coexist. In this sense, we
realize that there is a strong symmetry in SSR.}
\end{figure}

 \subsection{Flat space-time metric with the background frame $S_V$}

From Eq.(23), we obtain 

\begin{equation}
ds^2=\frac{1}{\left(1-\frac{V^2}{v^2}\right)}g_{\mu\nu}dx^{\mu}dx^{\nu},
\end{equation}
where we have $ds^{*2}=\theta^2 ds^2=(1-\alpha^2)ds^2=(1-V^2/v^2)ds^2=g_{\mu\nu}dx^{\mu}dx^{\nu}$ (Eq.(23)). 

Eq.(30) is written as

\begin{equation}
ds^2=\Theta g_{\mu\nu}dx^{\mu}dx^{\nu},
\end{equation}
where $\Theta$ is a function of speed $v$ given with respect to the background frame $S_V$, namely:

\begin{equation}
\Theta=\Theta(v)=\frac{1}{\left(1-\frac{V^2}{v^2}\right)}, 
\end{equation}
where we see that $\Theta=\theta^{-2}$.

The presence of the ultra-referential $S_V$ deforms the Minkowsky metric (Eq.(31)) and works like a uniform background field that
fills the whole flat space-time as a perfect fluid, playing the role of a kind of de-Sitter (dS) space-time ($\Lambda>0$)\cite{16}. 

The function $\Theta$ can be understood as being a scale factor that increases for very large wavelengths (cosmological scales)
governed by vacuum (dS), that is to say for much lower energies ($v\rightarrow V$) where we have $\Theta\rightarrow\infty$. Thus, the
factor $\Theta(=\theta^{-2})$ breaks strongly the invariance of $ds$ only for very large distances governed by vacuum of the 
ultra-referential $S_V$, leading to the cosmological anti-gravity governed by the tiny positive value of the cosmological constant
(section 6). In this regime of vacuum-$S_V$ ($v\rightarrow V$ or $\Theta\rightarrow\infty$), the interval $ds$ diverges. 

On the other hand, we have $\Theta\rightarrow 1$ for smaller scales of length, namely for higher energies ($v>>V$), 
where dS space-time approximates to the Minkowski metric as a special case, restoring the Lorentz symmetry and the invariance of $ds$. 

We realize that the presence of the background frame $S_V$ deforms the metric $g_{\mu\nu}$ by means of the scale factor $\Theta$, so that
we define a deformed flat metric $G_{\mu\nu}=\Theta g_{\mu\nu}$ that remains a diagonal matrix, but now having $\Theta$ in its diagonal,
namely: $\displaystyle G_{\mu\nu}=\Theta g_{\mu\nu}=
\begin{pmatrix}
\frac{1}{\left(1-\frac{V^2}{v^2}\right)}  & 0  & 0  & 0 \\
  0                           & -\frac{1}{\left(1-\frac{V^2}{v^2}\right)}  & 0   & 0 \\
  0                           & 0  & -\frac{1}{\left(1-\frac{V^2}{v^2}\right)}   & 0 \\
  0                           & 0           & 0     & -\frac{1}{\left(1-\frac{V^2}{v^2}\right)}
\end{pmatrix}$. Therefore, we simply write Eq.(31) as being $ds^2=G_{\mu\nu}dx^{\mu}dx^{\nu}$. If we make $v>>V$, this implies
$\Theta\rightarrow 1$ and, thus, we recover the Minkowsky metric $g_{\mu\nu}$. 

Now we are already able to conclude that there should be the same universal factor $\theta=\sqrt{det(\theta I)}(<1)$ that deforms all
the invariant scalars of SR as, for instance, the space-time interval, i.e., $\theta ds=ds^{*}$, and the proper time interval, i.e., 
$\theta d\tau=d\tau^{*}$, so that $\Delta s^{*}(=\theta\Delta s)$ and $\Delta\tau^{*}(=\theta\Delta\tau)$ are the invariant intervals 
in SSR.

In Section 4, we will see that the mass, energy and momentum are also deformed by the same factor $\theta$, i.e., 
$m_{(0,\alpha)}=\theta m_0$, $E=\theta mc^2=\theta\gamma m_0c^2=\Psi m_0c^2$ and $p=\theta\gamma m_0v=\Psi m_0v$. Thus, we already can conclude that all those
invariant quantities of SR and others like the rest mass (Section 4) are abandoned in SSR, since they are modified by the factor $\theta$
due to the presence of the ultra-referential $S_V$ connected to the own invariance of the minimum speed $V$. 

In sum, we should understand that, as the invariance of $c$ leads us to break down the newtonian invariance
of the improper time interval ($\Delta t=\Delta\tau$), by introducing the dilation of the improper time interval 
($\Delta t=\gamma\Delta\tau$), which still preserves the invariance of the proper time and the space-time interval, now with a further step 
towards a new invariance of a minimum speed $V$, we are led to break down such invariant quantities of SR, since the proper time interval
can also dilate by means of the new factor $\theta$ (Eq.(27)). So, the new invariant quantities in SSR are now the deformed intervals
$\Delta\tau^{*}$ and $\Delta s^{*}$. 

\subsection{The alternative mathematical structure for the transformation $\Lambda$}

Since the Lorentz matrix $L$ represents a rotation in spacetime ($det L=1$), it is known that $L$ can be alternatively
written in the form, namely: 

 \begin{equation}
\displaystyle L=
\begin{pmatrix}
\gamma  & -\beta\gamma \\
-\beta\gamma  & \gamma
\end{pmatrix}=
\displaystyle
\begin{pmatrix}
\cosh\phi  & -\sinh\phi \\
-\sinh\phi &  \cosh\phi
\end{pmatrix},
\end{equation}
where $\cosh\phi=\gamma=1/\sqrt{1-\beta^2}$, $\sinh\phi=\beta\gamma=\beta/\sqrt{1-\beta^2}$, $\tanh\phi=\beta$ and
$det L=(\cosh\phi)^2-(\sinh\phi)^2=1$. 

From Eq.(33), we obtain the following transformations: 

\begin{equation}
 x^{\prime}=(\cosh\phi)x-(\sinh\phi)ct\equiv\gamma(x-vt)
\end{equation}

and 

\begin{equation}
 ct^{\prime}=(\cosh\phi)ct-(\sinh\phi)x\equiv\gamma(ct-\beta x)
\end{equation}

Although we already know that the new matrix $\Lambda$ is not a rotation matrix, even so we will use the above hyperbolic representation 
for computing $\Lambda$, which will be given in function of those hyperbolic functions and, after, we will interpret the results. In order
to do that, we make a factoration of $\theta$ outside the matrix $\Lambda$, by writing it in the following way:  

\begin{equation}
\displaystyle\Lambda=\theta
\begin{pmatrix}
\gamma  & -\beta(1-\alpha)\gamma \\
-\beta(1-\alpha)\gamma  & \gamma
\end{pmatrix}=
\displaystyle\theta
\begin{pmatrix}
\cosh\phi  & -\sinh\phi(1-\alpha) \\
-\sinh\phi(1-\alpha) &  \cosh\phi
\end{pmatrix},
\end{equation}
where $\theta\gamma=\theta\cosh\phi=\Psi$, $\theta\beta\gamma=\theta\sinh\phi$ and 
$det\Lambda=\theta^2[(\cosh\phi)^2-(1-\alpha)^2(\sinh\phi)^2]\neq 1$. 

From Eq.(36), we obtain the following new transformations: 

\begin{equation}
x^{\prime}=\theta\left[(\cosh\phi)X-(\sinh\phi)ct+(\sinh\phi)\frac{V}{v}ct\right]\equiv\Psi(X-vt+Vt)
\end{equation}

and 

\begin{equation}
ct^{\prime}=\theta\left[(\cosh\phi)ct-(\sinh\phi)X+(\sinh\phi)\frac{V}{v}X\right]\equiv\Psi\left(ct-\frac{v}{c}X+\frac{V}{c}X\right)
\end{equation}

We realize that, if we make $V=0$, which implies $\theta=1$, the rest state is recovered, so that the background frame $S_V(X,Y,Z)$
(Fig.1) is eliminated and replaced by the galilean reference frame $S(x,y,z)$ at rest. Therefore, the new transformations in Eq.(37)
and Eq.(38) recover the well-known transformations of rotation given in Eq.(34) and Eq.(35). 

Since the angle $\phi$ in Eq.(37) and Eq.(38) cannot be understood as being simply a rotation that preserves the norm of the usual $4$-vector 
for the whole interval of speeds, i.e., $V<v<c$, let us deal with Eq.(37) and Eq.(38) by considering basically two regimes, namely: 

a) For $v>>V$, such that $\frac{V}{v}\approx 0$ (this is not necessarily a relativistic regime $v\approx c$), we find 
$\theta\approx 1$ and we can also neglect the terms $(\sinh\phi)\frac{V}{v}ct$ (Eq.(37)) and $(\sinh\phi)\frac{V}{v}X$ (Eq.(38))
with respect to the others, so that we recover the rotation regime within a good approximation, i.e., Eq.(34) and Eq.(35) begin to
take place. 

b) For $v\approx V$, such that $\frac{V}{v}\approx 1$ (this is a very low energy $E\approx 0$, which is obtained by making 
$v\rightarrow V$ in Eq.(42)), thus we can consider the following approximations: the second and third terms in the right member of both equations
(Eq.(37)) and Eq.(38)) are cancelled between themselves, so that we immediately obtain $x^{\prime}=\theta(\cosh\phi)X$ and 
$ct^{\prime}=\theta(\cosh\phi)ct$, with $\theta<<1$; however, as the rotations (boosts) do not exist close to the background
frame $S_V$, we still should make $\phi\rightarrow 0$ ($\cosh\phi\rightarrow 1$) in such approximations, such that we finally get $x^{*\mu}=\theta Ix^{\mu}$,
which is exactly Eq.(21), where we have replaced the index $^{\prime}$ by the index $^{*}$, since the boosts do not make sense in this
new regime, where $\theta$ just plays the role of a scale factor that deforms the interval $ds$. 

\section{Energy and momentum with the presence of a minimum speed}

Let us firstly define the $4$-velocity in the presence of the background frame $S_V$ connected to the invariant minimum speed $V$, 
as follows:

\begin{equation}
 U^{\mu}=\left[\frac{v_{\alpha}\sqrt{1-\frac{V^2}{v^2}}}{c\sqrt{1-\frac{v^2}{c^2}}}~ , 
~\frac{\sqrt{1-\frac{V^2}{v^2}}}{\sqrt{1-\frac{v^2}{c^2}}}\right]=\left[U^{\alpha},U^{4}\right]
\end{equation}
where $\mu=1,2,3,4$ and $\alpha=1,2,3$. If $V\rightarrow 0$, we recover the well-known 4-velocity of SR. From Eq.(39), it is interesting 
to observe that the 4-velocity of SSR vanishes in the limit of $v\rightarrow V$ ($S_V$), i.e., $U^{\mu}=(0,0,0,0)$, whereas in SR, 
for $v=0$ we find $U^{\mu}=(1,0,0,0)$.

The $4$-momentum is
\begin{equation}
 p^{\mu}=m_0cU^{\mu},
   \end{equation}
being $U^{\mu}$ given in Eq.(39). So we find

\begin{equation}
 p^{\mu}=\left[\frac{m_0v_{\alpha}\sqrt{1-\frac{V^2}{v^2}}}{\sqrt{1-\frac{v^2}{c^2}}}~ , 
~ \frac{m_0c\sqrt{1-\frac{V^2}{v^2}}}{\sqrt{1-\frac{v^2}{c^2}}}\right]=\left[p^{\alpha},p^{4}\right],
\end{equation}
where $p^4=E/c$, such that

\begin{equation}
E=cp^4=mc^2=m_0c^2\frac{\sqrt{1-\frac{V^2}{v^2}}}{\sqrt{1-\frac{v^2}{c^2}}},
\end{equation}
where $E$ is the total energy of the particle with speed $v$ in relation to the background reference frame (ultra-referential $S_V$). 
From Eq.(42), we observe that, if $v\rightarrow c\Rightarrow E\rightarrow\infty$. If $v\rightarrow V\Rightarrow E\rightarrow 0$ and, if
$v=v_0=\sqrt{cV}\Rightarrow E=E_0=m_0c^2$ (proper energy in SSR), where we should stress that $m_0c^2$ requires a non-zero motion
 $v(=v_0)$ in relation to $S_V$. Figure 4 shows us the graph for the energy $E$. 

\begin{figure}
\begin{center}
\includegraphics[scale=0.7]{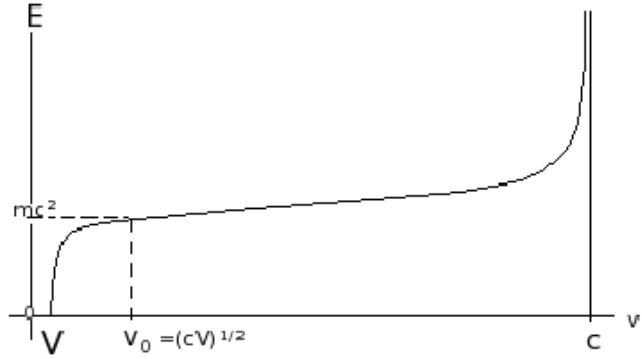}
\end{center}
\caption{$v_0=\sqrt{cV}$ is a speed such that we get the proper energy of the particle ($E_0=m_0c^2$) in SSR, where $\Psi_0=\Psi(v_0)=1$. 
For $v<<v_0$ or closer to $S_V$ ($v\rightarrow V$), a new relativistic correction on energy arises, so that $E\rightarrow 0$. On the other 
hand, for $v>>v_0$, being $v\rightarrow c$, so we find $E\rightarrow\infty$.}
\end{figure}

From Eq.(41) we also obtain the $3$(spatial)-momentum, namely:

\begin{equation}
\vec p = m_0\vec v\frac{\sqrt{1-\frac{V^2}{v^2}}}{\sqrt{1-\frac{v^2}{c^2}}},
\end{equation}
where $\vec v=(v_1,v_2,v_3)$.

From Eq.(41), performing the quantity $p^{\mu}p_{\mu}$, we obtain the energy-momentum relation of SSR, as follows:

\begin{equation}
p^{\mu}p_{\mu}=\frac{E^2}{c^2}-\vec p^2=m_0^2c^2\left(1-\frac{V^2}{v^2}\right),
\end{equation}
where $\vec p^2=p_1^2+p_2^2+p_3^2$. 

From Eq.(44), we obtain

\begin{equation}
E^2=c^2p^2+m_0^2c^4\theta^2=c^2p^2+m_0^2c^4\left(1-\frac{V^2}{v^2}\right)
\end{equation}

In SR theory, that is represented by the Lorentz group, some elements are preserved under rotations, as for instance, the 
$4$-interval $ds^2(=g_{\mu\nu}dx^{\mu}dx^{\nu})$ and also the rest mass by means of the inner product $p^{\mu}p_{\mu}=m_0^2c^2$,
that is the dispersion relation, where the rest mass is conserved. This means that the rest condition and the rest mass are fundamental
in SR, since they are independent of the state of motion, i.e., we have the well-known indistinguishability of motion and rest. However,
the new dispersion relation given in Eq.(44) (or Eq.(45)) shows us that the rest condition does not exist, since now the mass depends on
its preferred state of motion $v$ with respect to the background frame connected to an invariant minimum speed $V$, i.e., the
ultra-referential $S_V$. This is the reason why we find the massive term as a function of $\alpha$, i.e., we get $m_0^2c^4(1-\alpha^2)$ 
in Eq.(44) and Eq.(45). 

In Eq.(44), when, $\alpha\rightarrow 1$ ($v\rightarrow V$), we find $p^{\mu}p_{\mu}\rightarrow 0$, however, we can never nullify 
$p^{\mu}p_{\mu}$, since the minimum speed $V$ is unattainable (see Section 5). 
 
In the present work, as we are focusing our attention on some dynamical implications of a minimum speed, let us leave a more detailed
development of the physical consequences of SSR in terms of field-theory actions to be explored elsewhere. However, here it would be
interesting to mention that the wave operator is covariant under the new transformations.  

In order to obtain the new dispersion relation in an alternative and simple way, being consistent with the result obtained from the 
formalism of $4$-momentum, we have to consider $p=\Psi m_0v$, $E=\Psi m_0 c^2$ and $E_0=m_0c^2$, such that, we first calculate
the quantity $c^2p^2$ and we obtain

\begin{equation}
c^2p^2=\frac{m_0^2c^2(v^2-V^2)}{\left(1-\frac{v^2}{c^2}\right)}
\end{equation}

On the other hand, we find

\begin{equation}
E^2-E^2_0=m^2c^4-m_0^2c^4=\frac{m_0^2c^4}{\left(1-\frac{v^2}{c^2}\right)}\left[\frac{v^2}{c^2}-\frac{V^2}{v^2}\right]
\end{equation}

It is easy to verify that, if we make $V=0$ in Eq.(46) and in Eq.(47), we recover the well-known dispersion relation of SR, i.e., 
$c^2p^2=E^2-E^2_0=m^2c^4-m_0^2c^4$. However, according to Eq.(46) and Eq.(47), we see that $(E^2-m_0^2c^4)\neq c^2p^2$. So, in order
to obtain a new energy-momentum relation (new dispersion relation) with the presence of the minimum speed $V$, we should compare 
Eq.(46) with Eq.(47) by introducing a certain correction function $A(v)$, so that now we write the following identity: 

\begin{equation}
E^2-m_0^2c^4=\frac{m_0^2c^4}{\left(1-\frac{v^2}{c^2}\right)}\left[\frac{v^2}{c^2}-\frac{V^2}{v^2}\right]\equiv
\frac{m_0^2c^2(v^2-V^2)}{\left(1-\frac{v^2}{c^2}\right)}+A(v)=c^2p^2+A(v),
\end{equation}
where $A(v)$ should be found in order to satisfy the identity in Eq.(48). After performing some calculations we find 
$A(v)=-m_0^2\alpha^2 c^4=-m_0^2c^4(V^2/v^2)$. Now by inserting $A(v)$ into Eq.(48), we finally obtain the relation, namely
$E^2-m_0^2c^4=c^2p^2+A(v)=c^2p^2-m_0^2c^4(V^2/v^2)$, from where we get

\begin{equation}
 E^2=c^2p^2+m_0^2c^4-m_0^2c^4\left(\frac{V^2}{v^2}\right)=c^2p^2+m_0^2c^4\left(1-\frac{V^2}{v^2}\right),
\end{equation}
that is the same relation in Eq.(45). Therefore, we can conclude that a certain massive term in this spacetime has always connection with
the state of motion with respect to the preferred frame-$S_V$, where, according to Eq.(49), we can write the effective mass as
$m_{(0,\alpha)}=\theta m_0=m_0\sqrt{1-V^2/v^2}$, which does not represent a rest mass $m_0$, since $v>V$. In view of this, we can also 
write the total energy, as follows: 

\begin{equation}
E=m_{(0,\alpha)}c^2 + K=\gamma m_{(0,\alpha)}c^2=\frac{\sqrt{1-\frac{V^2}{v^2}}}{\sqrt{1-\frac{v^2}{c^2}}}m_0c^2,
\end{equation}
where $K$ is the knetic energy and $E=\gamma m_{(0,\alpha)}c^2=\gamma\theta m_0c^2=\Psi m_0c^2$. So, from Eq.(50) we obtain $K$, namely: 

\begin{equation}
K= m_{(0,\alpha)}c^2(\gamma-1)=m_0c^2\sqrt{1-\frac{V^2}{v^2}}\left(\frac{1}{\sqrt{1-\frac{v^2}{c^2}}}- 1\right), 
\end{equation}

where $K\rightarrow 0$ if $v\rightarrow V$. If $V\rightarrow 0$ in Eq.(51), we recover the relativistic knetic energy, i.e.,
$K=m_0c^2(\gamma-1)$. 

Making an expansion in Eq.(51) and consider the approximation $v<<c$, we find

\begin{equation}
K=m_0c^2\sqrt{1-\frac{V^2}{v^2}}\left(1+\frac{v^2}{2c^2}+...-1\right)\approx\frac{1}{2}\left(m_0\sqrt{1-\frac{V^2}{v^2}}\right)v^2
=\frac{1}{2}m_{(0,\alpha)}v^2, 
\end{equation}
where $m_{(0,\alpha)}=m_0\theta(v)=m_0\sqrt{1-\frac{V^2}{v^2}}$.

Now, also making the approximation $v>>V$ ($\alpha\approx 0$) in Eq.(52), i.e., $m_{(0,\alpha)}\approx m_{(0,0)}=m_0$, we finally obtain
the approximation $V<<v<<c$, so that we simply find 

\begin{equation}
K\approx\frac{1}{2}m_0v^2,  
\end{equation}
which is the newtonian knetic energy, that is recovered only for intermediary speeds in such a spacetime with an invariant minimum 
speed.

The de-Broglie wavelength of a particle is due to its motion $v$ with respect to $S_V$, namely:

\begin{equation}
 \lambda=\frac{h}{p}=\frac{h}{\Psi m_0v}=\frac{h}{m_0v}\frac{\sqrt{1-\frac{v^2}{c^2}}}{\sqrt{1-\frac{V^2}{v^2}}},
 \end{equation}
from where we have used the momentum $p=\Psi m_0v=\theta\gamma m_0v$ given with respect to $S_V$ ((Eq.(43)).

If $v\rightarrow c\Rightarrow\lambda\rightarrow 0$ (spatial contraction) and $p\rightarrow\infty$. If $v\rightarrow V(S_V)\Rightarrow\lambda\rightarrow\infty$
(spatial dilation by breaking down Lorentz symmetry), which means that we have very large wavelengths. This leads to 
$\Theta\rightarrow\infty$ (see Eq.(30) and Eq.(31)) and $p\rightarrow 0$, since we can alternatively 
write $p=\theta\gamma m_0v=\Theta^{-1/2}\gamma m_0v$, where $\Theta=\theta^{-2}=1/(1-\alpha^2)$. 

\subsection{Transformations of momentum-energy in the presence of the ultra-referential $S_V$}
 
 By considering the quadri-vector of momentum-energy given in Eq.(41), we have $p^{\mu}=[p^{\alpha}, E/c]$. Since we already have
considered the motion in only one dimension (e.g: $x$), we obtain the vector $[p^{1}, E/c]$, where $p^{1}=p_x$.

Now, as we want to investigate how $p^{\mu}$ transforms in such a spacetime with the presence of the ultra-referential $S_V$, we have
to make those two transformations by using the matrix $\Lambda$ (Eq.(3)) and its inverse $\Lambda^{-1}$ (Eq.(6)). So, by first
considering $\Lambda$, we rewrite 

\begin{equation}
\displaystyle\Lambda=
\begin{pmatrix}
\Psi & -\beta (1-\alpha)\Psi \\
-\beta (1-\alpha)\Psi & \Psi
\end{pmatrix},
\end{equation}
such that the direct matricial transformation $p^{\prime\nu}=\Lambda^{\nu}_{\mu} p^{\mu}$ ($S_V\rightarrow S^{\prime}$) leads to the 
new momentum-energy transformations, as follows: 

\begin{equation} 
 p^{\prime}_x=\Psi\left[p_x-\frac{v(1-\alpha)E}{c^2}\right]=\Psi\left(p_x-\frac{vE}{c^2}+\frac{VE}{c^2}\right),  
\end{equation}
being $p^{\prime}_y=p_y$ and  $p^{\prime}_z=p_z$.

\begin{equation} 
 E^{\prime}=\Psi\left[E-v(1-\alpha)p_x\right]=\Psi\left(E-vp_x+Vp_x\right)
\end{equation}

We know that the inverse matrix (Eq.(6)) that transforms $S^{\prime}\rightarrow S_V$ is 

\begin{equation}
\displaystyle\Lambda^{-1}=
\begin{pmatrix}
\Psi^{\prime} & \beta (1-\alpha)\Psi^{\prime} \\
\beta (1-\alpha)\Psi^{\prime} & \Psi^{\prime}
\end{pmatrix},
\end{equation}
where we find $\Psi^{\prime}=\Psi^{-1}/[1-\beta^2(1-\alpha)^2]$. Thus, the inverse matricial transformation
$p^{\nu}=\Lambda^{-1\nu}_{\mu} p^{\prime\mu}$ ($S^{\prime}\rightarrow S_V$) leads to the following momentum-energy transformations, 
namely: 

\begin{equation} 
 p_x=\Psi^{\prime}\left[p^{\prime}_x+\frac{v(1-\alpha)E^{\prime}}{c^2}\right]=
\Psi^{\prime}\left(p^{\prime}_x+\frac{vE^{\prime}}{c^2}-\frac{VE^{\prime}}{c^2}\right),  
\end{equation}
being $p_y=p^{\prime}_y$ and  $p_z=p^{\prime}_z$.

\begin{equation} 
 E=\Psi^{\prime}\left[E^{\prime}+v(1-\alpha)p^{\prime}_x\right]=\Psi^{\prime}\left(E^{\prime}+vp^{\prime}_x-Vp^{\prime}_x\right)
\end{equation}

The Lorentz transformations of the energy-momentum $p^{\mu}$ are simply recovered if we make $V=0$ (Fig.5).  

\begin{figure}
\begin{center}
\includegraphics[scale=0.9]{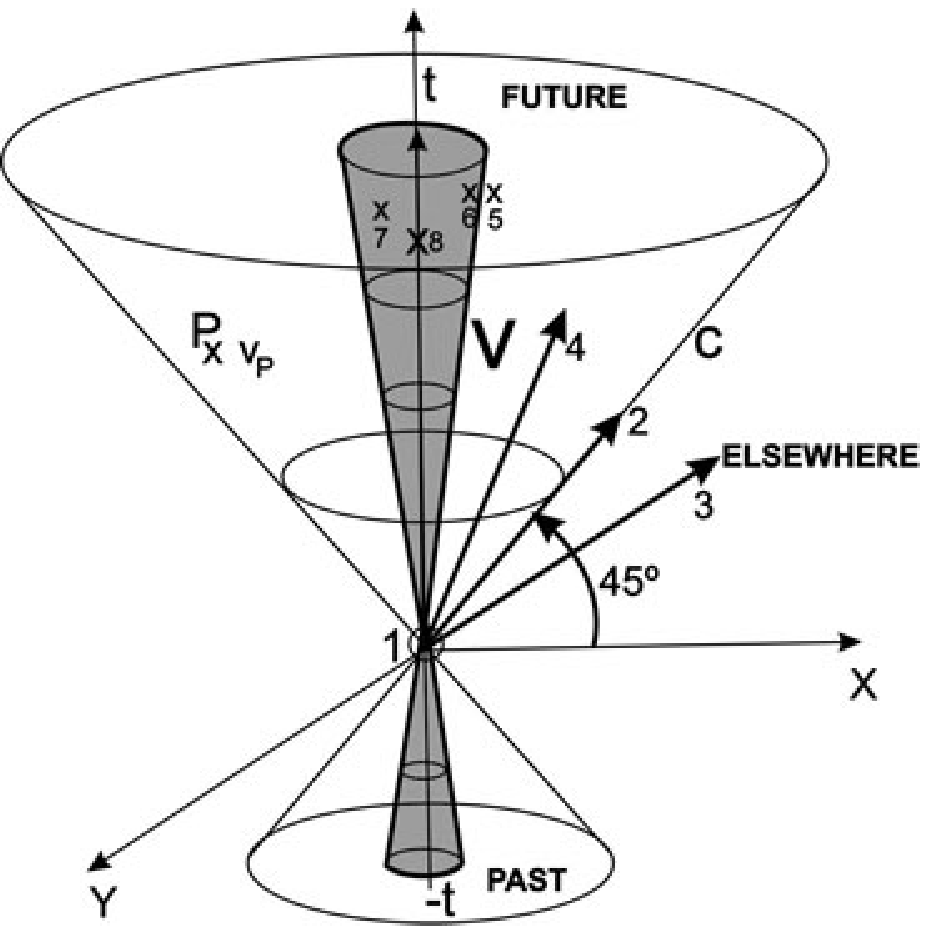}
\end{center}
\caption{The external and internal conical surfaces represent respectively the speed of light $c$ and the unattainable minimum speed $V$,
 where $V$ is represented by the dashed line, namely a definitely prohibited boundary for any particle. For a point $P$ in the world line
 of a particle, in the interior of the two conical surfaces, we obtain a corresponding internal conical surface, 
 such that we must have $V<v_p\leq c$. See a detailed explanation for this figure in:
 http://www.worldscientific.com/worldscinet/ijmpd?journalTabs=read :``On the electrodynamics of moving particles in a quasi flat
 spacetime with Lorentz violation and its cosmological implications'' }
\end{figure}

\section{Power of an applied force: the energy barrier of a minimum speed connected to the vacuum energy}

Let us consider a force applied to a particle, in the same direction of its motion. More general cases where the force is not 
necessarily parallel to velocity will be treated elsewhere. In our specific case ($\vec F||\vec v$), the relativistic power
$P_{ow}(=vdp/dt)$ of SSR is given as follows:

\begin{equation}
P_{ow}=v\frac{d}{dt}\left[m_0v\left(1-\frac{V^2}{v^2}\right)^{\frac{1}{2}}\left(1-\frac{v^2}{c^2}\right)^{-\frac{1}{2}}\right],
\end{equation}
where we have used the momentum $p$ given in Eq.(43).

After performing the calculations in Eq.(61), we find

\begin{equation}
 P_{ow}=\left[\frac{\left(1-\frac{V^2}{v^2}\right)^{\frac{1}{2}}}{\left(1-\frac{v^2}{c^2}\right)^{\frac{3}{2}}}
+\frac{V^2}{v^2\left(1-\frac{v^2}{c^2}\right)^{\frac{1}{2}}\left(1-\frac{V^2}{v^2}\right)^{\frac{1}{2}}}\right]
\frac{dE_k}{dt},
\end{equation}
where $E_k=\frac{1}{2}m_0v^2$.

If we make $V\rightarrow 0$ and $c\rightarrow\infty$ in Eq.(62), we simply recover the power obtained in newtonian mechanics, namely
$P_{ow}=dE_k/dt$. Now, if we just consider $V\rightarrow 0$ in Eq.(62), we recover the well-known relativistic power of SR, namely
$P_{ow}=(1-v^2/c^2)^{-3/2}dE_k/dt$. We notice that such a relativistic power tends to infinite ($P_{ow}\rightarrow\infty$) in the
limit $v\rightarrow c$. We explain this result as an effect of the drastic increase of an effective inertial mass close to $c$, namely
$m_{eff}=m_0(1-v^2/c^2)^{k^{\prime\prime}}$, where $k^{\prime\prime}=-3/2$. We must stress that such an effective inertial mass is the
response to an applied force parallel to the motion according to Newton second law, and it increases faster than the relativistic 
mass $m=m_r=m_0(1-v^2/c^2)^{-1/2}$.

  The effective inertial mass $m_ {eff}$ we have obtained is a longitudinal mass $m_L$, i.e., it is a response to the force applied in the
 direction of motion. In SR, for the case where the force is perpendicular to velocity, we can show that the transversal mass increases like the relativistic
 mass, i.e., $m=m_T=m_0(1-v^2/c^2)^{-1/2}$, which differs from the longitudinal mass $m_L=m_0(1-v^2/c^2)^{-3/2}$.  So, in this sense, there is 
anisotropy of the effective inertial mass to be also investigated in more details by SSR in a further work.

 The mysterious discrepancy between the relativistic mass $m$ ($m_r$) and the longitudinal inertial mass $m_L$ from Newton second law
 (Eq.(62)) is a
 controversial issue\cite{17}\cite{18}\cite{19}\cite{20}\cite{21}\cite{22}. Actually the newtonian notion about inertia
 as the resistance to acceleration
($m_L$) is not compatible with the relativistic dynamics ($m_r$) in the sense that we generally cannot consider $\vec F=m_{r}\vec a$. The dynamics of SSR
 aims to give us a new interpretation for the inertia of the newtonian point of view in order to make it compatible with the relativistic
 mass. This compatibility
 will be possible just due to the influence of the background field that couples to the particle and ``dresses" its relativistic mass in
 order to generate an
 effective (dressed) mass in accordance with the newtonian notion about inertia from Eq.(61) and Eq.(62). This issue will be
 clarified in this section.

 From Eq.(62), it is important to observe that, when we are closer to $V$, there emerges a completely new result (correction)
 for power, namely:

\begin{equation}
P_{ow}\approx\left(1-\frac{V^2}{v^2}\right)^{-\frac{1}{2}}\frac{d}{dt}\left(\frac{1}{2}m_0v^2\right),
\end{equation}
given in the approximation $v\approx V$. So, we notice that $P_{ow}\rightarrow\infty$ when $v\approx V$. We can also make the limit
$v\rightarrow V$ for the general case (Eq.(62)) and so we obtain an infinite power ($P_{ow}\rightarrow\infty$). Such a new relativistic effect deserves
the following very important comment:  Although we are in the limit of very low energies close to $V$, where the energy of the particle ($mc^2$) tends
to zero according to the approximation $E=mc^2\approx m_0c^2(1-V^2/v^2)^{k}$ with $k=1/2$ (e.g.: make the approximation $v\approx V$ 
in Eq.(42)), on the
other hand the power given in Eq.(63) shows us that there is an effective inertial mass that increases to infinite in the limit $v\rightarrow V$, that is to
say, from Eq.(63) we get the effective mass $m_{eff}\approx m_0(1-V^2/v^2)^{k^{\prime}}$, where $k^{\prime}=-1/2$. Therefore, from a dynamical point of view,
 the negative exponent $k^{\prime}$ ($=-1/2$) for the power at very low speeds (Eq.(63)) is responsible for the inferior barrier of the
 minimum speed $V$, as well as the exponent $k^{\prime\prime}=-3/2$ of the well-known relativistic power is responsible for the top barrier
 of the speed of light $c$ according to Newton second law. Actually, due to the drastic increase of $m_{eff}$ of a particle moving
 closer to $S_V$, leading to its strong coupling to the vacuum field in the background frame $S_V$, thus, in view of this, the dynamics 
 of SSR states that it is impossible to decelerate a subatomic particle until reaching the rest. 

 In order to see clearly both exponents $k^{\prime}=-1/2$ (inferior inertial barrier $V$) and $k^{\prime\prime}=-3/2$ (top inertial barrier
 $c$), let us write the general formula of power (Eq.(62)) in the following alternative way after some algebraic manipulations on it,
 namely:

\begin{equation}
P_{ow}=\left(1-\frac{V^2}{v^2}\right)^{k^{\prime}}\left(1-\frac{v^2}{c^2}\right)^{k^{\prime\prime}}\left(1-\frac{V^2}{c^2}\right)
\frac{dE_k}{dt},
\end{equation}
where $k^{\prime}=-1/2$ and $k^{\prime\prime}=-3/2$. Now it is easy to see that, if $v\approx V$ or even $v<<c$, Eq.(64) recovers the
approximation in Eq.(63). As $V<<c$, the ratio $V^2/c^2$ in Eq.(64) is a very small dimensionless constant. So it could be neglected.

From Eq.(64) we get the effective inertial mass $m_{eff}$ of SSR, namely:

\begin{equation}
m_{eff}=m_0\left(1-\frac{V^2}{v^2}\right)^{-\frac{1}{2}}\left(1-\frac{v^2}{c^2}\right)^{-\frac{3}{2}}\left(1-\frac{V^2}{c^2}\right)
\end{equation}

We must stress that $m_{eff}$ in Eq.(65) is a longitudinal mass $m_L$. The problem of mass anisotropy will be treated elsewhere, where we
will intend to show that, just for the approximation $v\approx V$, the effective inertial mass becomes practically isotropic, that is to
say $m_L\approx m_T\approx m_0\left(1-\frac{V^2}{v^2}\right)^{-1/2}$. This important result will show us the isotropic aspect of the
 vacuum-$S_V$, 
so that the inferior barrier $V$ has the same behavior of response ($k^{\prime}=-1/2$) of a force applied at any direction
in the space, namely for any angle between the applied force and the velocity of the particle.

 We must point out the fact that $m_{eff}$ has nothing to do with the ``relativistic mass" (relativistic energy $E$ in Eq.(42)) in the
 sense that
$m_{eff}$ is dynamically responsible for both barriers $V$ and $c$. The discrepancy between the ``relativistic mass" (energy $mc^2$ of the particle) and
such an effective inertial mass ($m_{eff}$) can be interpreted under SSR theory, as follows: $m_{eff}$ is a dressed inertial mass given 
in response to the presence of the vacuum-$S_V$ that works like a kind of ``fluid" in which the particle $m_0$ is immersed, 
while the ``relativistic mass" in SSR (Eq.(42)) works like a
 bare inertial mass in the sense that it is not considered to be under the dynamical influence of the ``fluid" connected to the
 vacuum-$S_V$. That is the reason
 why the exponent $k=1/2$ in Eq.(42) cannot be used to explain the existence of an infinite barrier at $V$, namely the vacuum-$S_V$
 barrier is governed by
the exponent $k^{\prime}=-1/2$ as shown in Eq.(63), Eq.(64) and Eq.(65), which prevents $v_*(=v-V)\leq 0$.

 The difference betweeen the dressed (effective) mass and the relativistic (bare) mass, i.e., $m_{eff}-m$ represents an interactive
 increment of mass $\Delta m_{i}$ that has purely origin from the vacuum energy of $S_V$, mamely: 

\begin{equation}
\Delta m_{i}= m_0\left[\frac{\left(1-\frac{V^2}{c^2}\right)}{\left(1-\frac{V^2}{v^2}\right)^{\frac{1}{2}}
\left(1-\frac{v^2}{c^2}\right)^{\frac{3}{2}}}- \frac{\left(1-\frac{V^2}{v^2}\right)^{\frac{1}{2}}}{\left(1-\frac{v^2}{c^2}\right)^{\frac{1}{2}}}\right]
\end{equation}

We have $\Delta m_{i}=m_{eff}-m$, being $m_{eff}=m_{dressed}$ given in Eq.(65) and $m$ ($m_r$) given in Eq.(42).

 From Eq.(66), we consider the following special cases:

a) for $v\approx c$ we have

\begin{equation}
 \Delta m_{i}\approx m_0\left[\left(1-\frac{v^2}{c^2}\right)^{-\frac{3}{2}}-\left(1-\frac{v^2}{c^2}\right)^{-\frac{1}{2}}\right]
\end{equation}

As the effective inertial mass $m_{eff}$ ($m_L$) increases much faster than the bare (relativistic) mass $m$ ($m_r$) close to the speed $c$,
 there is an increment of inertial mass $\Delta m_i$ that dresses $m$ in direction of its motion and it tends to be infinite when $v\rightarrow c$,
i.e., $\Delta m_i\rightarrow\infty$.

b) for $V<<v<<c$ (newtonian or intermediary regime) we find $\Delta m_i\approx 0$, where we simply have $m_{eff}$ ($m_{dressed}$)$\approx m\approx m_0$.
 This is the classical approximation.

c) for $v\approx V$ (close to the vacuum-$S_V$ regime), we have the following approximation:

\begin{equation}
\Delta m_{i}=(m_{dressed}-m)\approx m_{dressed}\approx\frac{m_0}{\sqrt{1-\frac{V^2}{v^2}}},
\end{equation}
where $m\approx 0$ when $v\approx V$ (see Eq.(42)).

The approximation in Eq.(68) shows that the whole dressed mass has purely origin from the energy of vacuum-$S_V$, with $m_{dressed}$ being
the pure increment $\Delta m_{i}$, since the bare (relativistic) mass $m$ of the own particle
 almost vanishes in such a regime ($v\approx V$), and thus an inertial effect only due to the vacuum (``fluid")-$S_V$
remains. We see that $\Delta m_{i}\rightarrow\infty$ when $v\rightarrow V$. In other words, we can interpret this infinite barrier of vacuum-$S_V$ by
considering the particle to be strongly coupled to the background field-$S_V$ in all directions of the space. The isotropy of $m_{eff}$ in this regime will
be shown in detail elsewhere, being $m_{eff}=m_L=m_T\approx m_0(1-V^2/v^2)^{-1/2}$. In such a regime, the particle practically loses its locality
(``identity") in the sense that it is spread out isotropically in the whole space and it becomes strongly coupled to the vacuum field-$S_V$, leading to an
 infinite value of $\Delta m_{i}$. Such a divergence of the dressed mass has origin from the dilation factor $\Theta_v(\rightarrow\infty)$ for this regime
when $v\approx V$, so that we can rewrite Eq.(68) in the following way: $\Delta m_{i}\approx m_{dressed}\approx m_0\Theta(v)^{1/2}$. That is
essentially the dynamical explanation why the particle cannot reach the rest in SSR theory so that the background frame of 
the vacuum-$S_V$ becomes unattainable for any particle at quantum level. However, in the macroscopic (classical) level, the
minimum speed $V$ as well as the Planck constant $\hbar$ are negligible as a good approximation, such that the rest state 
is naturally recovered in spite of the subatomic particles that constitute a body at rest are always moving, since its temperature
can never reach the absolute zero, as well as their constituent subatomic particles can never reach $V$. 

 Figure 6 shows the graph for the longitudinal effective inertial mass $m_{eff}=m_L$ ($m_{dressed}$) as a function of the speed $v$ with
respect to the ultra-referential $S_V$. 

\begin{figure}
\begin{center}
\includegraphics[scale=0.9]{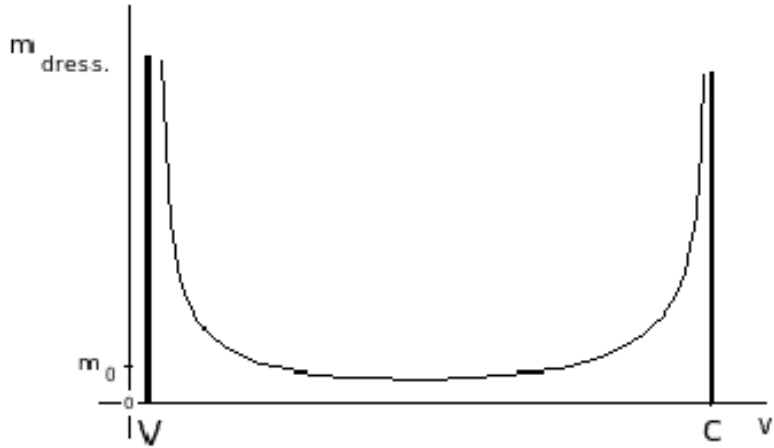}
\end{center}
\caption{The graph shows us two infinite barriers at $V$ and $c$, providing an aspect of symmetry of SSR. The first barrier ($V$) is exclusively due
 to the vacuum-$S_V$, being interpreted as a barrier of pure vacuum energy. In this regime we have the following approximations:
 $m_{eff}=m_{dressed}\approx\Delta m_{i}\approx m_0(1-V^2/v^2)^{-1/2}$ and  $m_r\approx m_0(1-V^2/v^2)^{1/2}$ (see Fig.4), so that
 $m_{dressed}\rightarrow\infty$
 and $m=m_r=m_{bare}\rightarrow 0$ when $v\rightarrow V$. The second barrier ($c$) is a sum (mixture) of two contributions, namely the own
 bare
 (relativistic) mass $m$ that increases with the factor $\gamma=(1-v^2/c^2)^{-1/2}$ (see Fig.4) plus the interactive increment $\Delta
 m_{i}$ due to the
 vacuum energy-$S_V$, so that $m_{dressed}=m_L=m+\Delta m_{i}\approx m_0(1-v^2/c^2)^{-3/2}$. This is a longitudinal effect. For the
 transversal effect,
  $\Delta  m_{i}=0$ since we get $m_T=m$. This result will be shown elsewhere.}
\end{figure}

  Now it is important to notice that a particle moving in one spatial dimension ($x$) goes only to right or to left, since the unattainable
 minimum limit of speed $V$ prevents it to reach the rest in view of the rapid increase of its dressed mass (Eq.(68)). So we cannot stop 
 the motion and return in the
 same spatial dimension $x$. On the other hand, in a complementary and symmetric way to $V$, the limit $c$, which represents the temporal aspect of the space-time,
  prevents to stop the marching of the time ($v_t=0$), and so avoiding to come back to the past. In short, we perceive that the basic
 ingredient of
 the space-time structure in SSR, namely the $(1+1)D$ space-time, presents $x$ and $t$ in equal-footing in the sense that both of them 
 are irreversible once the
 particle is moving only to right or to left. Such an equal-footing ``$xt$" in SSR theory does not occurs in SR theory since we can stop 
 the spatial motion in SR ($v_x=0$) and after come back in $x$, but not in the time
 $t$. However, if we take into account more than one spatial dimension in SSR theory, at least two spatial dimensions ($xy$), thus 
 the particle could return by moving in
 the additional dimension(s) $y$ (and/or $z$). So, SSR theory is able to provide the reason why we must have more than one ($1$) spatial
  dimension for representing movement
  in reality $(3+1)D$, although we could have only one ($1$) spatial dimension just as a good practical approximation for the 
  case of classical space-time as in SR theory (e.g.:a ball moving in a rectilinear path). 

\section{Cosmological implications}

\subsection{Energy-momentum tensor in the presence of the ultra-referential-$S_V$}

  Let us rewrite the 4-velocity (Eq.(39)) in the following alternative way: 
   \begin{equation}
 U^{\mu}=\left[\frac{\sqrt{1-\frac{V^2}{v^2}}}{\sqrt{1-\frac{v^2}{c^2}}}~ , ~
\frac{v_{\alpha}\sqrt{1-\frac{V^2}{v^2}}}{c\sqrt{1-\frac{v^2}{c^2}}}\right]=\left[U^0, U^{\alpha}\right], 
   \end{equation}
where now we have $\mu=0,1,2,3$ and $\alpha=1,2,3$. If $V\rightarrow 0$, we recover the $4$-velocity of SR.

The well-known energy-momentum tensor to deal with perfect fluid is of the form
   \begin{equation}
  T^{\mu\nu}=(p+\epsilon)U^{\mu}U^{\nu} - pg^{\mu\nu},
   \end{equation}
where $U^{\mu}$ is given in Eq.(69). $p$ represents a pressure and $\epsilon$ an energy density.

From Eq.(69) and Eq.(70), by performing the new component $T^{00}$, we obtain
   \begin{equation}
  T^{00}=\frac{\epsilon(1-\frac{V^2}{v^2})+p(\frac{v^2}{c^2}-\frac{V^2}{v^2})}{(1-\frac{v^2}{c^2})}
   \end{equation}

 If $V\rightarrow 0$, we recover the old component $T^{00}$.

Now, as we are interested only in obtaining $T^{00}$ in absence of matter, i.e.,
 the vacuum limit connected to the ultra-referential $S_V$, we perform the limit of Eq.(71) as follows:
 \begin{equation}
 lim_{v\rightarrow V} T^{00}= T^{00}_{vacuum}=\frac{p(\frac{V^2}{c^2}-1)}{(1-\frac{V^2}{c^2})}= -p.
  \end{equation}

 From Eq.(71), we notice that the term $\epsilon\gamma^2(1-V^2/v^2)$ for representing matter naturally vanishes
 in the limit of vacuum-$S_V$ ($v\rightarrow V$), and therefore just the contribution of vacuum prevails. As we always
 must have $T^{00}>0$, we get $p<0$ in Eq.(72). This implies a negative pressure for vacuum energy density of the
ultra-referential $S_V$. So we verify that a negative pressure emerges naturally from such a new tensor in the limit of $S_V$.

  We can obtain $T^{\mu\nu}_{vacuum}$ by calculating the limit of vacuum-$S_V$ for Eq.(70), by considering Eq.(69), as follows:

 \begin{equation}
 T^{\mu\nu}_{vacuum}= lim_{v\rightarrow V}T^{\mu\nu}= -pg^{\mu\nu},
 \end{equation}
 where we conclude that $\epsilon=-p$. In Eq.(69), we see that the new $4$-velocity vanishes in the limit of the vacuum-$S_V$ ($v\rightarrow V$),
 namely $U^{\mu}_{vac.}=(0,0)$. So, $T^{\mu\nu}_{vac.}$ is in fact a diagonalized tensor as we hope to be. The vacuum-$S_V$ that is
 inherent to such a space-time with an invariant minimum speed works like a {\it sui generis} fluid in equilibrium with negative pressure, 
leading to a cosmological anti-gravity, i.e., the invariant minimum speed connected to a universal background field in the preferred
 frame $S_V$ leads naturally to the well-known equation of state of the cosmological constant $p=w\epsilon$, with $w=-1$\cite{31}. 

\subsection{The cosmological constant $\Lambda$ and the vacuum energy density $\rho$}

 The well-known relation\cite{31} between the cosmological constant $\Lambda$ and the vacuum energy density $\rho_{(\Lambda)}$ is

\begin{equation}
\rho_{(\Lambda)}=\frac{\Lambda c^2}{8\pi G}
\end{equation}

Let us consider a spherical universe with Hubble radius filled by a uniform vacuum energy density. On the surface of such a
sphere (frontier of the observable universe), the bodies (galaxies) experience an accelerated expansion (anti-gravity) due to
the whole ``dark mass (energy)" of vacuum inside the sphere. So we could think that each galaxy is a proof body interacting with that big sphere
like in the simple case of two bodies interaction. However, we need to show that there is an anti-gravitational interaction between the 
ordinary proof mass $m$ and the big sphere with a ``dark mass" of vacuum ($M_{\Lambda}$), but let us first start from the well-known 
simple model of a massive proof particle $m_0$ that escapes from a classical gravitational potential $\phi$
on the surface of a big sphere of matter, namely $E=m_0c^2(1-v^2/c^2)^{-1/2}\equiv m_0c^2(1+\phi/c^2)$, where $E$ is its relativistic
energy. Here the interval of escape velocity $0\leq v<c$ is associated with the interval of potential $0\leq\phi<\infty$, where we 
stipulate $\phi>0$ to be the attractive (classical) gravitational potential. 

Now we can show that the influence of the background field (vacuum energy inside the sphere) connected to the ultra-referential
$S_V$ (see Eq.(72)) leads to a strong repulsive (negative) gravitational potential ($\phi<<0$) for very low energies ($E\rightarrow 0$). 
In order to see this non-classical aspect of gravitation\cite{23}, we use Eq.(42) just taking into account the new approximation given
for very low energies $(v(\approx V)<<c)$, as follows: 

\begin{equation}
E\approx\theta m_0c^2=m_0c^2\sqrt{1-\frac{V^2}{v^2}}\equiv m_0c^2\left(1+\frac{\phi}{c^2}\right),
\end{equation}
where $\phi<0$ (repulsive). For $v\rightarrow V$, this implies $E\rightarrow 0$, which leads to $\phi\rightarrow -c^2$.
So, the non-classical (most repulsive) minimum potential $\phi(V)(=-c^2)$
connected to vacuum-$S_V$ is responsible for the cosmological anti-gravity (see also Eq.(72) and Eq.(73)). We interpret this result
assuming that only an exotic ``particle" of the vacuum energy at $S_V$ could escape from the anti-gravity ($\phi=-c^2$) generated by the 
vacuum energy inside the sphere (consider $v=V$ in Eq.(75)). Therefore, ordinary bodies like galaxies and any matter on the surface 
of such a sphere cannot escape from its anti-gravity, being accelerated far away.

 According to Eq.(75), we should note that such an exotic ``particle" of vacuum (at $S_V$) has an infinite mass $m$ since we should 
consider $v=V$ ($\theta=0$) in order to have a finite value of $E$, other than the photon ($v=c$), that is a massless particle
(see Eq.(42)). So we conclude that an exotic ``particle" of vacuum works like a counterparty of the photon, namely an infinitely
massive boson.

  We consider that the most negative (repulsive) potential ($\phi=-c^2$ for $v=V$, in Eq.(75)) is related to the cosmological
constant (vacuum energy density), since we have shown in Eq.(72) and Eq.(73) that the background reference frame $S_V$ plays the role of 
the vacuum energy density with a negative pressure, working like the cosmological constant $\Lambda$ ($p=-\epsilon=-\rho_{(\Lambda)}$).
So we write

\begin{equation}
\phi_{\Lambda}=\phi(\Lambda)=\phi(V)=-c^2
\end{equation}

Let us consider the simple model of spherical universe with a radius $R_u$, being filled by a uniform vacuum energy
density $\rho_{(\Lambda)}$, so that the total vacuum energy inside the sphere $E_{\Lambda}=\rho_{(\Lambda)}V_u=-pV_u=M_{\Lambda}c^2$.
$V_u$ is its volume and $M_{\Lambda}$ is the total dark mass associated with the dark energy for $\Lambda$ 
(vacuum energy: $w=-1$\cite{31}). Therefore the repulsive gravitational potential on the surface of such a sphere is

\begin{equation}
\phi_{\Lambda}=-\frac{GM_{\Lambda}}{R_u}=-\frac{G\rho_{(\Lambda)}V_u}{R_uc^2}=\frac{4\pi GpR_u^2}{3c^2},
\end{equation}
where $p=-\rho_{(\Lambda)}$, with $w=-1$\cite{31}.

 By introducing Eq.(74) into Eq.(77), we find

\begin{equation}
\phi_{\Lambda}=\phi(\Lambda)=-\frac{\Lambda R_u^2}{6}
\end{equation}

Finally, by comparing Eq.(78) with Eq.(76), we extract

\begin{equation}
\Lambda=\frac{6c^2}{R_u^2},
\end{equation}
where $\Lambda S_u=24\pi c^2$, $S_u=4\pi R_u^2$.

And, by comparing Eq.(77) with Eq.(76), we have

\begin{equation}
\rho_{(\Lambda)}=-p=\frac{3c^4}{4\pi G R_u^2},
\end{equation}
where $\rho_{(\Lambda)} S_u=3c^4/G$. We can verify that Eq.(80) and Eq.(79) satisfy Eq.(74).

 In Eq.(79), $\Lambda$ is a kind of {\it cosmological scalar field}, extending the old concept of
Einstein about the cosmological constant for stationary universe. From Eq.(79), by considering the Hubble radius, with
$R_{u}=R_{H_0}\sim 10^{26}$m, we obtain $\Lambda=\Lambda_0\sim (10^{17}m^2s^{-2}/10^{52}m^2)\sim 10^{-35}s^{-2}$.
To be more accurate, we know the age of the universe $T_0=13.7$ Gyr, being $R_{H_0}=cT_0\approx
1.3\times 10^{26}$m, which leads to $\Lambda_0\approx 3\times 10^{-35}s^{-2}$. It is interesting to notice that this tiny positive value
is in agreement with the observational data\cite{24}\cite{25}\cite{26}\cite{27}\cite{28}. The vacuum energy
density\cite{29}\cite{30} given in Eq.(80) for $R_{H_0}$ is $\rho_{(\Lambda_0)}\approx
2\times 10^{-29}g/cm^{3}$, which is also in agreement with observations. For the scale of the Planck length, where
$R_{u}=l_P=(G\hbar/c^3)^{1/2}$, from Eq.(79) we find $\Lambda=\Lambda_P=6c^5/G\hbar\sim 10^{87}s^{-2}$, and from Eq.(80), 
$\rho_{(\Lambda)}=\rho_{(\Lambda_P)}=T^{00}_{vac.P}=\Lambda_P c^2/8\pi G=3c^7/4\pi G^2\hbar\sim 10^{113}J/m^3
(=3c^4/4\pi l_P^2G\sim 10^{43}kgf/S_P\sim 10^{108}atm\sim 10^{93}g/cm^3)$. So, just at that past time, $\Lambda_P$ or
$\rho_{(\Lambda_P)}$ played the role of an inflationary vacuum field with 122 orders of magnitude\cite{31} beyond 
the ones ($\Lambda_0$ and $\rho_{(\Lambda_0)}$) for the present time.

It must be stressed that our assumption for obtaining the tiny value of $\Lambda_0$ starts from first principles related to
a new symmetry in spacetime, i.e., we have introduced the idea of a background reference frame $S_V$ for representing the vacuum energy 
connected to an invariant minimum speed $V$, leading to the cosmological constant.

Here it should be also emphasized that both the cosmological constant and the minimum speed have non-zero values ​​due to the same cause,
which is essentially the existence of a fundamental state of vacuum with non-zero (very low) energy density, given by 
the potential $-c^2$ in Eq.(76). This means that the non-zero value of the cosmological constant $\Lambda_0(\sim 10^{-35}s^{-2})$ has origin in the
fact that there must be also a non-zero value of a minimum speed $V$ at subatomic level; however this does not mean that the value
of $V$ should be obtained directly as a function of $\Lambda$, since we just know that $V(>0)$ and $\Lambda(>0)$ are different aspects 
of the same reality, i.e., the existence of a non-null energy density of vacuum. In any way, a deeper investigation of the origin of
the minimum speed $V$ by obtaining its value should be important in order to clarify further this question. 

\section{Conclusion and prospects}

The very high values ​​obtained for the cosmological constant and the vacuum energy density by means of the quantum field theory for the quantum 
vacuum have a discrepancy of about 120 orders of magnitude beyond their observational values. This puzzle is well-known as the
{\it ``Cosmological Constant Problem"}\cite{31}. The idea of ​​an invariant minimum speed $V$ connected to a background field for the
ultra-referential $S_V$, within an extended structure of spacetime by breaking down Lorentz symmetry, has led to low values of the vacuum 
energy density and the cosmological constant in agreement with observational results of Perlmutter, Schmidt and Riess. 

After investigating the origin of the minimum speed $V$ and a possible connection between $V$ and the zero-point energy of the quantum
mechanics (the uncertainty principle), we should thoroughly explore many interesting consequences of SSR and its new dispersion relation in
quantum field theories (QFT), since the existence of a mimimum speed for lower energies with the same status of the speed of light for 
higher energies leads to a new metric for describing such deformed (symmetric) spacetime, allowing us to build a modified QFT, where
the Lorentz symmetry is broken down. This kind of metric $\Theta(v)g_{\mu\nu}$ in Eq.(31) is a special case of metric that has already
been explored in the literature and it seems to lead to the Finsler's geometry, namely a Finslerian space with a metric that depends on
the position and also the velocity, i.e., $G_{\mu\nu}(x,\dot{x})$\cite{32}\cite{33}\cite{34}.\\

{\noindent\bf  Acknowledgements}

 I am specially grateful to Prof. Jonas Durval Cremasco and Juliano S. Gonschorowski for interesting discussions. I thank Carlos Magno
 Leiras, C\'assio Guilherme Reis, A. C. Amaro de Faria Jr., Alisson Xavier, Em\'ilio C. M. Guerra, G. Vicentini and Paulo R. Souza Coelho
 for their comprehension of this work's significance.\\

\end{document}